\documentclass[twocolumn]{aastex631}

\usepackage[T1]{fontenc}
\usepackage{xspace}
\usepackage{multirow}
\usepackage{amsmath}
\usepackage{lipsum}
\usepackage{soul}
\usepackage[normalem]{ulem}
\usepackage{float}
\usepackage{comment}
\usepackage{graphicx}
\usepackage{soul}
\usepackage[normalem]{ulem}
\newcommand\soutpars[1]{\let\helpcmd\sout\parhelp#1\par\relax\relax}
 \usepackage{cancel}

\newcommand{\E}[1]{\ensuremath{\times 10^{#1}} }

\newcommand{\msol}{\ensuremath{M_{\odot}}\xspace}
\newcommand{\rxte}{\textit{RXTE}\xspace}

\newcommand{\nicer}{{NICER}\xspace}

\newcommand{\source}{4U~1820--30\xspace}

\newcommand{\fa}{$f_a$~}

\received{2024 November}
\accepted{2025 April}

\shorttitle{{Spectral Features observed during X-ray bursts and after a long burst from 4U~1820--30}}
\shortauthors{Jaisawal et al.}
\graphicspath{{./}{figures/}}
\begin{document}
\title{On the Origin of Spectral Features Observed during Thermonuclear X-ray Bursts and in the Aftermath Emission of a Long Burst from 4U~1820--30}
\author[0000-0002-6789-2723]{Gaurava K. Jaisawal}\email{gaurava@space.dtu.dk}\affiliation{DTU Space, Technical University of Denmark, Elektrovej 327-328, DK-2800 Lyngby, Denmark}

\author[0000-0002-4397-8370]{J{\'e}r{\^o}me Chenevez}
\affiliation{DTU Space, Technical University of Denmark, Elektrovej 327-328, DK-2800 Lyngby, Denmark}

\author[0000-0001-7681-5845]{Tod E. Strohmayer} 
 \affil{Astrophysics Science Division and Joint Space-Science Institute, NASA's Goddard Space Flight Center, Greenbelt, MD 20771, USA}

\author[0000-0003-1674-4859]{Hendrik Schatz}
\affiliation{Department of Physics and Astronomy, Michigan State University, East Lansing, MI 48824, USA}
\affiliation{Facility for Rare Isotope Beams, Michigan State University, East Lansing, Michigan 48824, USA}
\affiliation{Joint Institute for Nuclear Astrophysics-Center for the Evolution of the Elements (JINA-CEE), USA}

\author[0000-0002-4363-1756]{J.~J.~M.~in 't Zand}
\affil{SRON Netherlands Institute for Space Research, Sorbonnelaan 2, 3584 CA Utrecht, The Netherlands}

\author[0000-0002-3531-9842]{Tolga G\"uver}
\affiliation{Istanbul University, Science Faculty, Department of Astronomy and Space Sciences, Beyaz\i t, 34119, \.Istanbul, T\"urkiye}
\affiliation{Istanbul University Observatory Research and Application Center, Istanbul University 34119, \.Istanbul T\"urkiye}  

\author[0000-0002-3422-0074]{Diego Altamirano}
\affiliation{School of Physics and Astronomy, University of Southampton, Southampton SO17 1BJ, UK}

\author{Zaven Arzoumanian} 
\affiliation{Astrophysics Science Division, NASA's Goddard Space Flight Center, Greenbelt, MD 20771, USA}

\author[0000-0001-7115-2819]{Keith C. Gendreau} 
\affiliation{Astrophysics Science Division, NASA's Goddard Space Flight Center, Greenbelt, MD 20771, USA}

\begin{abstract}
We study 15 thermonuclear X-ray bursts from 4U~1820--30 observed with the Neutron Star Interior Composition Explorer (\nicer). We find evidence of a narrow emission line at 1.0~keV and three absorption lines at 1.7, 3.0, and 3.75~keV, primarily around the photospheric radius expansion phase of most bursts. The 1.0~keV emission line remains constant, while the absorption features, attributed to wind-ejected species, are stable but show slight energy shifts, likely due to combined effects of Doppler and gravitational redshifts. We also examine with \nicer the ``aftermath'' of a long X-ray burst (a candidate superburst observed by MAXI) on 2021 August 23 and 24. The aftermath emission recovers within half a day from a flux depression. During this recovery phase, we detect two emission lines at 0.7 and 1~keV, along with three absorption lines whose energies decreased to 1.57, 2.64, and 3.64~keV. Given the nature of the helium white-dwarf companion, these absorption lines during the aftermath may originate from an accretion flow, but only if the accretion environment is significantly contaminated by nuclear ashes from the superburst. This provides evidence of temporary metal enhancement in the accreted material due to strong wind loss. Moreover, we suggest that the absorption features observed during the short X-ray bursts and in the superburst aftermath share a common origin in heavy nuclear ashes enriched with elements like Si, Ar, Ca, or Ti, either from the burst wind or from an accretion flow contaminated by the burst wind.

\end{abstract}


\section{Introduction} \label{sec:intro}

Type-I X-ray bursts occur due to unstable thermonuclear burning in the surface layers of weakly magnetized neutron stars (NSs) in low-mass X-ray binary systems. The NS accretes mainly hydrogen and/or helium from its Roche-lobe filling companion star, which is usually a low-mass star ($<3\msol$) or a white dwarf \citep{Lewin1993SSRv...62..223L, Strohmayer2006csxs.book..113S, 2021ASSL..461..209G}. These bursts last from tens to hundreds of seconds, depending on the ignition depth and the composition of the burning material \citep{Bildsten2000AIPC..522..359B}. Observed burst profiles show a rapid rise followed by an exponential decay, representing the cooling of the NS surface. Short type-I bursts, lasting up to tens of seconds, are attributed to the triple-$\alpha$ process in a helium-rich fuel, while longer bursts up to few hundred seconds are expected from hydrogen-rich or mixed H/He fuels \citep{Fujimoto1981ApJ...247..267F, Bildsten2000AIPC..522..359B, Cumming2003ApJ...595.1077C}. The continuum burst emission is often described by a thermal blackbody component with a temperature ranging from 1 to 3~keV. A lower temperature of around 0.2~keV has been detected at the peak of bursts with an expansion radius from $\ge$200 to 1000~km \citep{Zand2010A&A...520A..81I, Keek2018ApJ...856L..37K, Yu2023arXiv231216420Y, Jaisawal2024}.

In the case of highly energetic bursts, the NS photosphere can expand by a few tens to hundreds of km, leading to photospheric radius expansion (PRE;  \citealt{Ebisuzaki1983PASJ...35...17E, Cumming2003ApJ...595.1077C, Kuulkers2003}). This phenomenon is observed at (super-) Eddington luminosities when the radiation pressure momentarily overcomes the gravitational pull of the NS. 
Strong radiatively driven winds from PRE bursts spread the heavy nuclear ashes synthesized during thermonuclear processes into the NS photosphere \citep{Weinberg2006ApJ...639.1018W, Yu2018ApJ...863...53Y, 2023A&A...678A.156H}. The signature of these elements can be detected as absorption lines or edges in X-ray burst spectra \citep{Weinberg2006ApJ...639.1018W}.
Understanding the effect of the gravitational redshift, {\it z}, on spectral features can enable the measurement of the NS compactness, the ratio between mass and radius parameters, through the relation 1+{\it z} = $(1 - 2GM/c^2R)^{-1/2}$. Observationally, the gravitational redshift can be estimated by measuring a change in the rest-frame energy of a line feature due to the gravitational potential of the NS. The parameter $z$ is represented as $(E_0 - E)/ E$, where $E_0$ and $E$ correspond to the centroid line energies of the feature in the rest-frame and at the NS surface, respectively.  Evidence of spectral lines or edges from X-ray bursts has been found from  
4U~0614+091 \citep{Zand2010A&A...520A..81I}, 4U~1722--30 \citep{Zand2010A&A...520A..81I}, \source \citep{Strohmayer1997ApJ...487L..77S, Zand2010A&A...520A..81I, Strohmayer2019ApJ...878L..27S}, IGR~J17062--6143 \citep{Degenaar2013ApJ...767L..37D}, HETE~J1900.1--2455 \citep{Kajava2017MNRAS.464L...6K}, and GRS~1747--312 \citep{Li2018ApJ...866...53L} that have allowed measurement of NS compactness in some of these sources. 

We study powerful X-ray bursts from \source in search of spectral features from heavy-element ashes with \nicer.  This system is an ultra-compact X-ray binary situated in the globular cluster NGC~6624 at a distance of 8.4~kpc in the Sagittarius constellation \citep{Grindlay1976ApJ...205L.127G, Stella1987ApJ...312L..17S, Valenti2004MNRAS.351.1204V}. Based on its short orbital period of 11.4~minutes \citep{Stella1987ApJ...312L..17S}, the NS in the system is expected to accrete from a hydrogen-deficient dwarf star or a helium white-dwarf with a mass of 0.06--0.07\msol  \citep{Rappaport1987ApJ...322..842R}. The persistent, accretion-driven X-ray emission from \source is usually observed in a high-soft state, where soft-thermal photons from the accretion disk dominate the continuum emission \citep{Priedhorsky1984ApJ...284L..17P, Farrell2009MNRAS.393..139F}. However, the source quasi-periodically transits to a low-hard state about every 170--176 days \citep{Chou2001ApJ...563..934C, Strohmayer2002ApJ...566.1045S}, where a non-thermal component dominates the persistent emission. Short but powerful X-ray bursts are detected in the low-hard state of \source at a recurrence time of 2--4 hours  \citep{Grindlay1976ApJ...205L.127G, Chou2001ApJ...563..934C, Galloway2008ApJS..179..360G, Zand2012A&A...547A..47I}. 
\nicer observed 15 X-ray bursts from the source between 2017 August and 2021 May. All these are PRE bursts with a maximum expansion radius ranging between 50 and 1000~km. The spectral properties of the burst emission and its influence on the accretion environment have been reported in \citet{Yu2023arXiv231216420Y} and \citet{Jaisawal2024}. A candidate burst oscillation at 716~Hz was detected from one burst at 2.9$\sigma$ significance level based on Monte Carlo simulations. This suggests that \source likely hosts (one of) the fastest rotating NS among the known X-ray binary systems \citep{Jaisawal2024}. 

Previous studies of persistent accretion emission from \source have shown the presence of a faint, relativistically smeared iron emission line around 6.97~keV \citep{Cackett2010ApJ...720..205C} or at 6.6~keV \citep{Marino2023MNRAS.525.2366M}. The iron emission feature could also be modeled as two absorption edges at 6.93 and 7.67~keV \citep{Mondal2016MNRAS.461.1917M}.  No other significant absorption or emission lines were found in the 2--10~keV persistent spectrum with XMM-Newton \citep{Costantini2012A&A...539A..32C}, NICER, and NuSTAR \citep{Marino2023MNRAS.525.2366M}. Below 1~keV, absorption features from O~VIII ($\approx$0.6~keV) and the Ne~IX complex ($\approx$0.9~keV), attributed to the interstellar medium, have been observed \citep{Costantini2012A&A...539A..32C}.

Spectral features at 1.0 (emission), 1.7, and 3.0 keV (both in absorption) were discovered in two pairs of bursts observed with \nicer from \source in 2017 August \citep{Strohmayer2019ApJ...878L..27S}. This finding was attributed to gravitational and Doppler effects on the spectral features originating from burst-driven winds during PRE bursts. This paper reports the evolution of spectral features around 1, 1.7, 3, and 3.75~keV observed from a larger sample of thermonuclear X-ray bursts from \source between 2017 and 2021 with \nicer. We also detected discrete spectral signatures around similar energies in the aftermath emission of a long burst observed with \nicer on 2021 August 23 and 24. Our study provides a comprehensive overview of these spectral features observed during different events from \source. We further explore their potential origin in radiation-driven burst winds and/or accretion scenarios, including the impact of the long burst on the accretion environment. This paper is organized as follows: Section~2 reports the observation and data analysis, followed by spectral analysis of the X-ray bursts (Section~3) and the aftermath emission of the long burst in 2021 August (Section~4). The discussion and conclusion from our study are presented in Sections~5 and 6, respectively.

\begin{figure}[]
\centering
\includegraphics[height=4.8in, width=3.3in, angle=0]{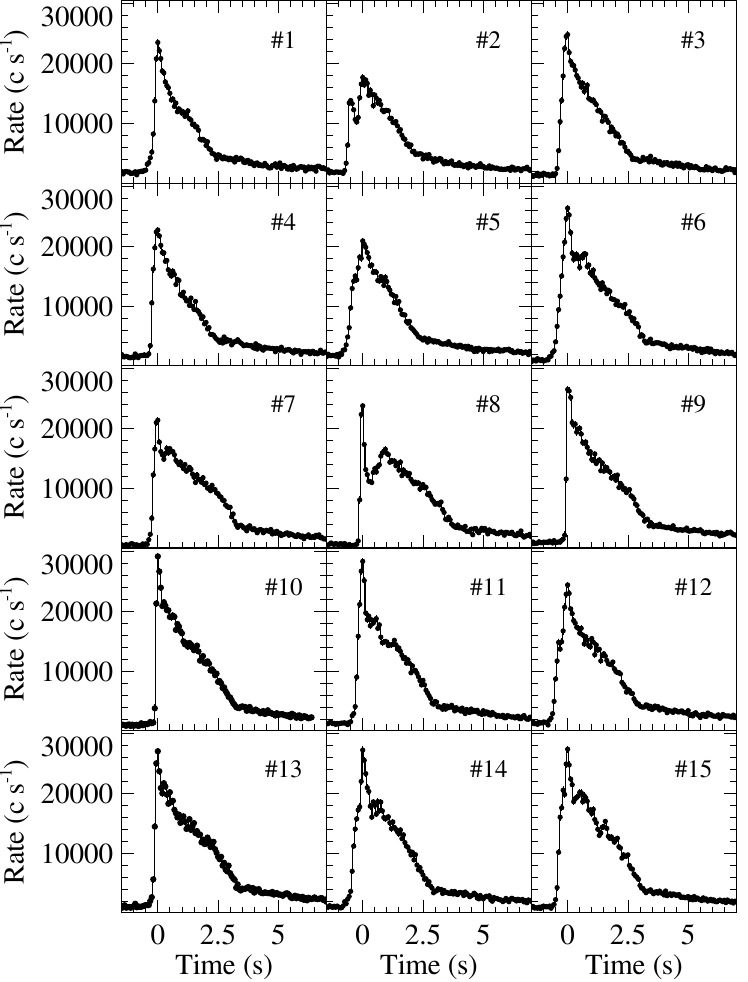}
\caption{All 15 thermonuclear X-ray bursts from \source observed with \nicer between 2017 August and 2021 May. Each burst light curve is shown at a time resolution of 1/16~s between 0.4--12 keV range, with the peak at time zero.   } 
\label{burst_lc}
\end{figure}


\begin{table}[]
\caption{ \nicer observations of X-ray bursts from \source.}
\begin{tabular}{lcccc}
\hline 
Burst   &ObsID  &Onset date  &Burst peak \\
Number    &       &(MJD)      &rate$^\dagger$ (c~s$^{-1}$) \\
\hline
1   &1050300108 &57994.37034   &21712$\pm$613 \\
2   &1050300108 &57994.46090   &15936$\pm$532 \\
3   &1050300109 &57995.22207   &23376$\pm$630  \\
4   &1050300109 &57995.33811   &21088$\pm$602 \\
5   &1050300109 &57995.60251   &19184$\pm$578 \\
6   &2050300104 &58641.28949   &25328$\pm$649 \\
7   &2050300108 &58646.83376   &20768$\pm$584 \\
8   &2050300110 &58648.32026   &23152$\pm$617 \\
9   &2050300115 &58655.08555   &25456$\pm$651 \\
10  &2050300119 &58660.30662   &28144$\pm$682 \\
11  &2050300119 &58660.76490   &27088$\pm$672 \\
12  &2050300120 &58661.77954   &23072$\pm$624  \\
13  &2050300122 &58663.97779   &25872$\pm$657 \\
14  &2050300124 &58665.58749   &26128$\pm$660 \\
15  &4680010101 &59336.60807   &26400$\pm$662 \\ 
\hline
\end{tabular}
\label{tab:bursts}\\
\footnotesize{$^\dagger$: Pre-burst count rates subtracted. }\\
\end{table}

\begin{table*}[]
\centering
\caption{Spectral parameters around the peak (segment I) of 15 thermonuclear X-ray bursts from \source described with the variable persistent emission method and power-law approach without emission and absorption lines. The bursts with similar parameters are represented in pairs. 
}
\begin{tabular}{lccccc | cccccc}
\hline
 &  \multicolumn{5}{c|}{{\it Variable persistent emission model}} & \multicolumn{6}{c}{{\it Power-law approach}}  \\
\hline
Burst         &kT$\rm_{BB}$ &Radius$\rm_{BB}$  &\fa  &Flux$\rm_T$     &$\chi^2_\nu$ (dof)  &kT$\rm_{BB}$ &Radius$\rm_{BB}$  &$\Gamma$  &Norm$_\Gamma$   &Flux$\rm_T$     &$\chi^2_\nu$ (dof)\\
        &(keV)         &(km)             &     &($10^{-8}$)  &  &(keV)         &(km)        &      &     &($10^{-8}$) \\
\hline
Pair-1 \\
\#4	&0.72$\pm$0.03		&54$\pm$5		&9.2$\pm$0.2		&8.3$\pm$0.1	&1.36 (366)    &0.65$\pm$0.03		&74$\pm$5		&1.75$\pm$0.04	&6.6$\pm$0.2 &8.9$\pm$0.2	&1.35 (365) \\
\#5	&0.67$\pm$0.03		&56$\pm$6		&8.6$\pm$0.2	&7.8$\pm$0.1	&1.12 (367)   &0.64$\pm$0.03		&74$\pm$5		&1.74$\pm$0.04	&6.3$\pm$0.2 &8.5$\pm$0.2	&1.12 (366)\\
\\
Pair-2	\\																						
\#1	&0.54$\pm$0.02		&97$\pm$7		&7.2$\pm$0.3		&7.3$\pm$0.1	&1.29 (359)     &0.57$\pm$0.02		&101$\pm$6		&1.74$\pm$0.04	&5.5$\pm$0.2 &8.0$\pm$0.2	&1.28 (358)\\
\#3	&0.59$\pm$0.02		&91$\pm$6		&9.6$\pm$0.3		&7.9$\pm$0.1	&1.24 (352)    &0.58$\pm$0.02		&105$\pm$5		&1.89$\pm$0.05	&6.6$\pm$0.2 &8.5$\pm$0.2	&1.21 (351)\\

\\												
Pair-3	\\											
\#11	&0.50$\pm$0.01		&152$\pm$5		&9.2$\pm$0.3		&7.5$\pm$0.1	&1.34 (335)    &0.53$\pm$0.01		&145$\pm$5		&2.04$\pm$0.05	&5.0$\pm$0.2 &7.4$\pm$0.1	&1.13 (334)\\
\#13	&0.48$\pm$0.01		&166$\pm$5		&8.8$\pm$0.4		&7.2$\pm$0.1	&1.98 (317)     &0.54$\pm$0.01		&151$\pm$4		&2.4$\pm$0.1	&4.0$\pm$0.3 &7.2$\pm$0.1	&1.41 (316)\\

\\												
Pair-4	\\
\#14	&0.50$\pm$0.01		&157$\pm$5		&8.0$\pm$0.3		&6.9$\pm$0.1	&1.32 (327)    &0.52$\pm$0.01		&156$\pm$5		&1.82$\pm$0.05	&3.5$\pm$0.2 &6.5$\pm$0.1	&1.17 (326)\\
\#15	&0.50$\pm$0.01		&164$\pm$5		&9.6$\pm$0.4		&6.7$\pm$0.1	&1.64 (327)    &0.54$\pm$0.01		&150$\pm$4		&2.10$\pm$0.07	&4.4$\pm$0.3 &6.9$\pm$0.1	&1.33 (326)\\
\\																						
Pair-5 \\
\#6	&0.45$\pm$0.01		&197$\pm$5		&7.8$\pm$0.3		&5.9$\pm$0.1	&1.98 (303)   &0.51$\pm$0.01		&164$\pm$4		&2.4$\pm$0.1	&4.0$\pm$0.2 &6.7$\pm$0.2	&1.36 (302) \\
\#9	&0.47$\pm$0.01		&175$\pm$5		&8$\pm$0.4		&6.2$\pm$0.1	&1.67 (315)  &0.50$\pm$0.01		&164$\pm$5		&2.06$\pm$0.06	&3.7$\pm$0.2 &6.2$\pm$0.1	&1.44 (314) \\

\\
\hline
\#2	&0.99$\pm$0.1		&31$\pm$4		&6.9$\pm$0.2		&7.2$\pm$0.1	&1.15 (377)   &0.72$\pm$0.05		&51$\pm$6		&1.51$\pm$0.04	&5.4$\pm$0.2 &8.9$\pm$0.3	&1.15 (376) \\

\#7	&0.40$\pm$0.01		&255$\pm$6		&6.7$\pm$0.4		&4.4$\pm$0.1	&1.99 (274)   &0.45$\pm$0.01		&189$\pm$6		&2.58$\pm$0.09	&3.1$\pm$0.2 &5.7$\pm$0.2	&1.35 (273)\\
\#8	&0.28		&468		&3.9		&-	&2.53 (268)   &0.34$\pm$0.01		&273$\pm$12		&2.85$\pm$0.08	&3.2$\pm$0.2 &6.1$\pm$0.3	&1.60 (267)\\

\#10	&0.42$\pm$0.01		&246$\pm$6		&6.3$\pm$0.3		&5.7$\pm$0.1	&1.84 (289)    &0.46$\pm$0.01		&203$\pm$5		&2.41$\pm$0.09	&3.6$\pm$0.2 &6.6$\pm$0.1	&1.28 (288)\\
\#12	&0.56$\pm$0.01		&106$\pm$5		&9.6$\pm$0.3		&7.6$\pm$0.1	&1.38 (342)   &0.60$\pm$0.02		&109$\pm$4		&2.05$\pm$0.06	&5.5$\pm$0.2 &7.6$\pm$0.1	&1.18 (341)\\
\hline
\end{tabular}
\label{tab:burst-group}\\
\footnotesize{
Note: Here, kT$\rm_{BB}$ and Radius$\rm_{BB}$ are the blackbody temperature and corresponding radius, respectively. \fa is the scaling factor. Norm$_\Gamma$ stands for the normalization of the power-law component with a photon index of $\Gamma$ in the unit of keV~cm$^{-2}$~s$^{-1}$ at 1~keV. Total model flux (Flux$\rm_T$) is in the unit of erg~s$^{-1}$~cm$^{-2}$ in the 0.1-100 keV range.}\\
\end{table*}

\section{Observations and Data analysis} \label{sec:obs}
The \nicer X-ray Timing Instrument (XTI; \citealt{Gendreau2012}) is a non-imaging soft X-ray telescope, attached to the International Space Station. It comprises 56 co-aligned concentrator optics, each coupled with a silicon-drift detector sensitive to 0.2--12~keV X-ray photons \citep{Prigozhin2012}. \nicer has good spectral resolution of 100~eV and excellent timing sensitivity of 100~ns (rms). The peak effective area of the instrument is 1900~cm$^2$ at an energy of 1.5~keV from its 52 functioning detector units.

\nicer has monitored \source since its launch in 2017 June. A total of 15 thermonuclear bursts have been observed in data sets accumulated mainly between 2017 August and 2021 May from 11 ObsIDs, as summarized in Table~\ref{tab:bursts}. 
In our analysis, \nicer data were processed utilizing the \texttt{nicerl2} pipeline within the \texttt{HEASoft} v6.30 software. Filtering criteria, including elevation angle from the Earth limb $\texttt{ELV}>15^\circ$, bright Earth limb angle $\texttt{BR\_EARTH}>30^\circ$, source angular offset of $\texttt{ANG\_DIST}<0\fdg015$; undershoot rate range of $\texttt{underonly\_range}=$~0--400, overshoot rate range of $\texttt{overonly\_range}=$~0--2, were applied with \texttt{nimaketime}. 

We generated the burst light curves and spectra from clean event data using the {\tt XSELECT} package. Light curves of all 15 bursts in the 0.4--12~keV range are presented in Figure~\ref{burst_lc} at a time resolution of 1/16~s. The bursts show peak count rates between 15,900 and 28,200~c~s$^{-1}$  (Table~\ref{tab:bursts}). 
See also \citet{Jaisawal2024} for more details on the observations, burst duration, pre-burst count rate, and other burst properties.
For the spectral analysis, the background spectrum was estimated using the {\tt nibackgen3C50}\footnote{\url{https://heasarc.gsfc.nasa.gov/docs/nicer/tools/nicer_bkg_est_tools.html}} tool \citep{Remillard2022AJ....163..130R}. The spectral response matrix and ancillary response files were produced using {\tt nicerrmf} and {\tt nicerarf} tools, respectively.  In this paper, we conduct a spectral analysis in the 0.4--10 keV energy range with NICER using the {\tt XSPEC} package \citep{Arnaud1996}.

\section{Burst Time-resolved Spectroscopy} \label{spec-sec1}

In this study, we perform time-resolved spectroscopy to search for possible spectral signatures in all 15 X-ray bursts from \source. 
Each burst is divided into five broad segments to cover successive time intervals from the peak of the burst light curve. This allows us to investigate gradual changes in the detected spectral features as the photosphere reaches different heights above the NS surface during the bursts, covering from PRE to cooling decay phases. 
The time intervals were chosen to be between --0.1 \& +0.6~s (segment~I), +0.6 \& +1.4~s (segment~II), +1.4 \& +2.3~s (segment~III), +2.3 \& +3.3~s (segment~IV), and +3.3 \& +4.4~s (segment~V) from the peak of each burst light curve, respectively.
Note that segment~I is very similar to the PRE phase intervals investigated by \citet{Strohmayer2019ApJ...878L..27S}.
Burst time-resolved spectroscopy on finer time scales has been reported by \citet{Jaisawal2024}, where the impact of burst emission on the accretion environment has been investigated using different modeling approaches. 

Before attempting the time-resolved spectroscopy of all 15 bursts, we extracted the persistent spectrum prior to each burst. The exposure time of each pre-burst spectrum ranged between 125 and 1000~s depending on the data available in the same \nicer orbit. We fitted each spectrum using an absorbed blackbody component with a Comptonization model ({\tt Comptt} in {\tt XSPEC}; \citealt{Titarchuk1994ApJ...434..570T}). The {\tt Tbabs} component in {\tt XSPEC} with {\tt wilm} abundances \citep{Wilms2000} and {\tt verner} cross-sections \citep{Verner1996ApJ...465..487V} is utilized to describe the photoelectric absorption in our model. The results from the pre-burst spectroscopy are the same as Table~3 of \citet{Jaisawal2024}. The unabsorbed pre-burst emission flux was found between (2.1---6.3)$\times$10$^{-9}$~erg~s$^{-1}$~cm$^{-2}$, in the 0.5--10 keV range, corresponding to 5 to 16\% of the Eddington luminosity of a pure helium accreting NS.  This calculation assumes the Eddington luminosity at 3.5$\times$10$^{38}$~erg~s$^{-1}$ \citep{Lewin1993SSRv...62..223L, Bult2019ApJ...885L...1B} and a distance of 8.4~kpc.

A variable persistent emission method (also known as the $f_a$-method; \citealt{Worpel2013}) was first considered to fit the burst time-resolved spectra in the 0.4--10~keV range, following \citet{Jaisawal2024}. This method can describe the spectral excess observed in addition to the main burst blackbody component, by scaling the pre-burst emission by a multiplicative factor. The observed spectral excess is thought to be due to increased accretion induced by the strong burst irradiation.
This modeling approach can fit the time-resolved spectra of each burst. However, we noticed the presence of excess residuals in general that give rise to a higher reduced chi-squared for strong bursts with a blackbody expansion radius of more than 150~km (see Table~\ref{tab:burst-group}). The residuals are mostly observed around the peak, even during burst spectroscopy at finer time scales with the variable persistent emission method and reflection-based model (see \citealt{Jaisawal2024}).  

Alternatively to the variable persistent emission method, a model adding a power-law component to the fixed pre-burst emission parameters, along with a blackbody component describing the burst emission, provides a better fit for the burst spectra and the observed excess. The model can be expressed as {\tt Tbabs*(bbodyrad + power-law + fixed-pre-burst-model)} in {\tt XSPEC}.  The parameters obtained from the variable persistent emission method and the power-law approach are presented in Table~\ref{tab:burst-group} for segment~I for comparison.  
The uncertainties on spectral parameters are given for 68\% confidence level in Table~\ref{tab:burst-group}.  We compare the fluxes obtained from the power-law component and the scaled persistent level from the variable persistent emission method. Depending on the burst, the power-law fluxes range from 45 to 75\% of the respective total model flux (as given in Table 2), and are usually 10 to 20\% lower than the scaled persistent fluxes derived using the variable persistent emission method. This flux reduction is anticipated because the power-law component does not include the contribution from the pre-burst level, which is usually accounted for by the scaling factor in the case of the variable persistent emission method. As a result, we conclude that the model including a power-law can also fit time-resolved spectra of bursts from \source.

From Table~\ref{tab:burst-group},  we observe a relatively softer photon index for bursts with large expansion radii, possibly representing the extent of the corona cooling or the effect of reflection from an extended part of the accretion disk. Moreover, the ad-hoc power-law-based model does better describe the excess above the main blackbody components than the variable persistent emission method, particularly for strong expansion bursts. Notably, improvements in the reduced chi-squared value arise from the continuum modeling and are mostly independent of observed residuals for line features. Thus, we consider the power-law-based model as the best model for examining spectral features at different emission phases of these bursts.


\begin{figure}[]
\centering
\includegraphics[height=2.7in, width=3.28in, angle=0]{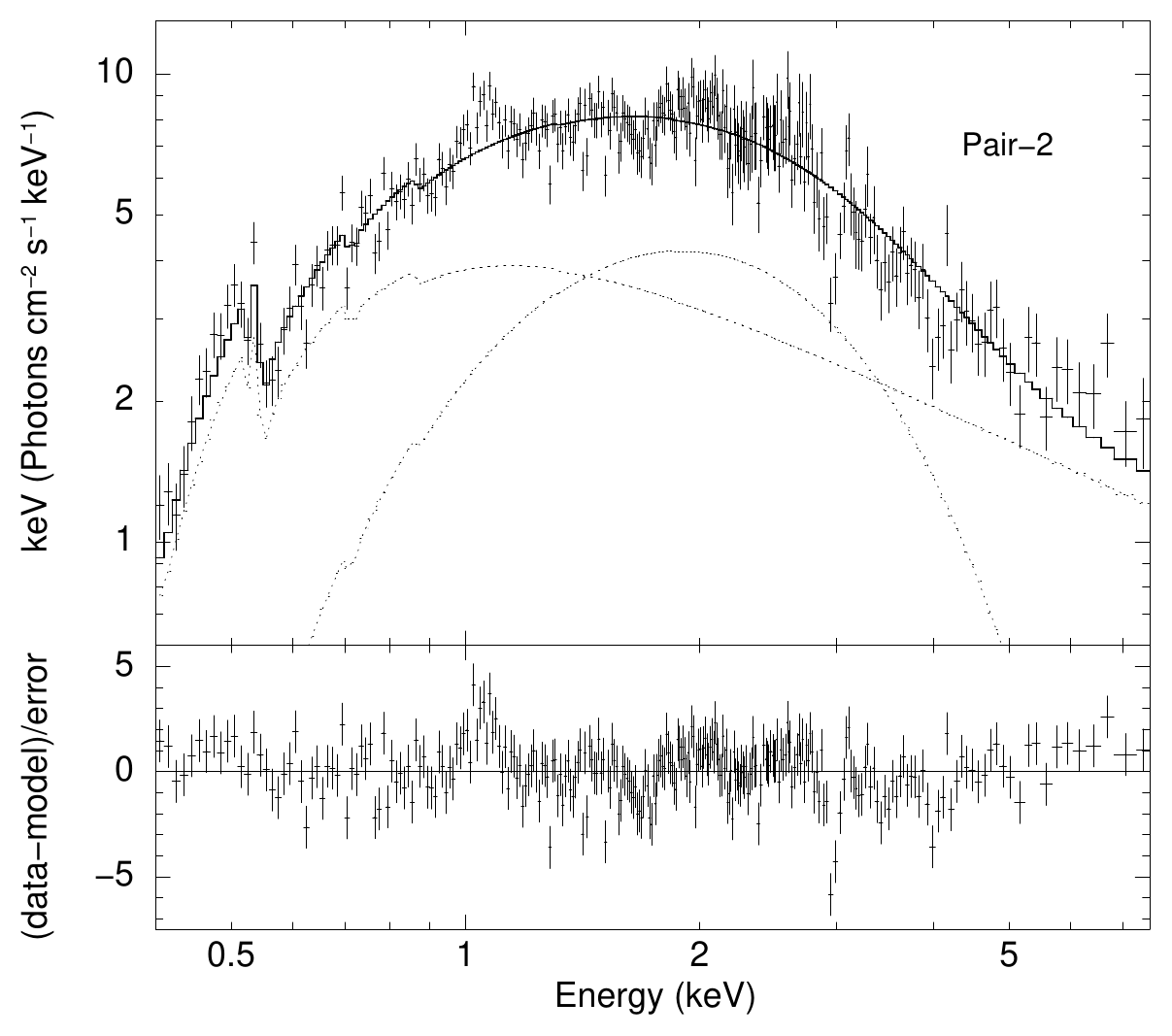}   
\includegraphics[height=2.7in, width=3.28in, angle=0]{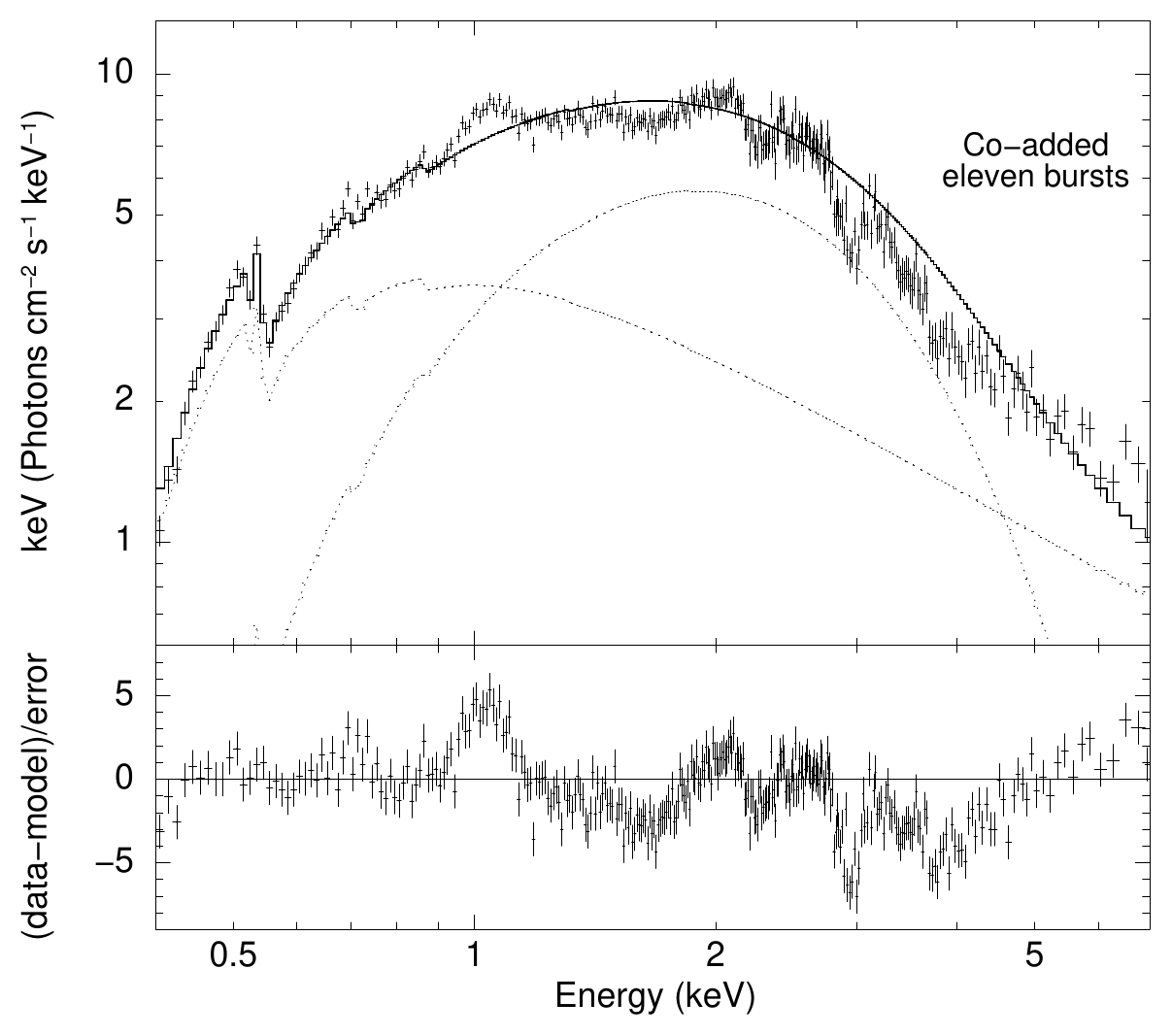}
\caption{The energy spectrum and the corresponding spectral residuals after fitting the co-added burst peak spectra from Pair-2 (top), using the power-law modeling approach.  The bottom figure shows a combined spectrum, comprising all peak spectra from eleven bursts listed in five pairs and Burst~\#12.  The 1 keV (emission), 1.7 keV absorption, 3 keV absorption, and 3.75~keV absorption features are evident in these spectra.  } 
\label{fig:group-spec}
\end{figure}


\begin{table*}[]
\centering
\scriptsize
\caption{Spectral parameters of observed line features in thermonuclear X-ray bursts of \source using a power-law modeling approach. Uncertainties are calculated at a 68\% confidence level. The co-added peak (segment I) spectra from five pairs are also given in the table.}
\begin{tabular}{lccccccccc}
\hline 
Burst        &Segment   &kT$\rm _{BB}$ &Radius$\rm _{BB}$           &Line Energy       &Line Width         &Line Norm     &Significance    &$\Delta\chi^2$ &Eqw. \\          
Number          &  &(keV)           &(km)           &(keV)            &(keV)          &         &($\sigma$)               &              &(eV)      \\
\hline
 			
1 & I	&	0.61	$\pm$	0.02	 &	95	$\pm$	5	&	1.05	$\pm$	0.01	&	$<$0.03	&	0.217	$\pm$	0.046	&	4.7	&	25	&	24 $\pm$ 9	\\																						
    & I	 &	0.61	$\pm$	0.02	 &	95	$\pm$	5	&	2.96	$\pm$	0.01	&	$<$0.05	&	-0.098	$\pm$	0.02	&	4.9	&	30	&	53 $\pm$ 10	\\
    & II	&	1.35	$\pm$	0.05	  &	22	$\pm$	1	&	1.03	$\pm$	0.03	&	0.07$\pm$0.02	&	0.182	$\pm$	0.06	&	3.1	&	18	&	30 $\pm$ 12	\\

				
3 & I	&	0.62	$\pm$	0.02	  &	96.6	$\pm$	4	&	1.04	$\pm$	0.01	&	$<$0.04	&	0.32	$\pm$	0.07	&	4.8	&	40	&	32 $\pm$ 6	\\
    & I	 &	0.62	$\pm$	0.02 &	96.6	$\pm$	4	&	2.97	$\pm$	0.01	&	$<$0.12	&	-0.096	$\pm$	0.02	&	4.7	&	26	&	50	$\pm$ 10 \\
    & II	&	1.02	$\pm$	0.06 &	29.8	$\pm$	2.6	&	1.04	$\pm$	0.01	&	$<$0.03	&	0.125	$\pm$	0.037	&	3.4	&	13	&	20 $\pm$ 7	\\
																			
4 & I	&	0.71	$\pm$	0.03 &	70	$\pm$	4	&	1.0	$\pm$	0.01	&	$<$0.03	&	0.20$\pm$0.05	&	4	&	19	&	21 $\pm$ 6	\\																							
    & I	&	0.71	$\pm$	0.03 &	70	$\pm$	4	&	2.85	$\pm$	0.02	&	0.08$\pm$0.03	&	-0.192	$\pm$	0.047	&	4	&	47	&	87 $\pm$ 17	\\
   &  II	&	1.2	$\pm$	0.1 &	19	$\pm$	3	&	1.05	$\pm$	0.06	&	$<$0.3	&	0.076	$\pm$	0.032	&	2.4	&	6	&	14 $\pm$ 10	\\
					
5      &  I	 &	0.67	$\pm$	0.03 &	73	$\pm$	5	&	1.03	$\pm$	0.04	&	$<$0.08	&	0.15	$\pm$	0.07	&	2.2	&	8	&	17 $\pm$ 9	\\
    & I	 &	0.67	$\pm$	0.03 &	73	$\pm$	5	&	1.62	$\pm$	0.01	&	$<$0.04	&	-0.086	$\pm$	0.026	&	3.3	&	12	&	17 $\pm$ 8	\\
    &  I	 &	0.67	$\pm$	0.03 &	73	$\pm$	5	&	3.06	$\pm$	0.05	&	0.11 $\pm$ 0.07	&	-0.12	$\pm$	0.05	&	2.3	&	14	&	66 $\pm$ 34	\\

6   & I &	0.53	$\pm$	0.01 &	155	$\pm$	5	&	1.02	$\pm$	0.03	&	0.07$\pm$0.04	&	0.41	$\pm$	0.133	&	3.1	&	38	&	39 $\pm$ 11\\
 & III	&	1.4	$\pm$	0.1 &	15.5	$\pm$	2.1	&	1.03	$\pm$	0.02	&	0.05$\pm$0.03	&	0.157	$\pm$	0.047	&	3.3	&	22	&	34 $\pm$ 8	\\
																				
7   & I	 &	0.47	$\pm$	0.01 &	172.5	$\pm$	9.3	&	1.03	$\pm$	0.01	&	0.03$\pm$0.02	&	0.397	$\pm$	0.096	&	4.1	&	31	&	41 $\pm$ 9	\\
    & I	 &	0.47	$\pm$	0.01 &	172.5	$\pm$	9.3	&	3.14	$\pm$	0.06	&	0.23$\pm$0.06	&	-0.237	$\pm$	0.067	&	3.4	&	44	&	284 $\pm$ 96	\\

8   & I	&	0.43	$\pm$	0.01 &	171	$\pm$	10	&	1.05	$\pm$	0.01	&	0.09	$\pm$	0.04	&	0.71	$\pm$	0.11	&	6.3	&	128	&	94 $\pm$ 32	\\
    & I	&	0.43	$\pm$	0.01 &	171	$\pm$	10	&	3.00	$\pm$	0.02	&	$<$0.05	&	-0.048	$\pm$	0.007	&	6.7	&	69	&	103	$\pm$ 47 \\

    & II	&	0.58	$\pm$	0.01 &	110	$\pm$	4	&	1.06	$\pm$	0.01	&	$<$	0.06	&	0.264	$\pm$	0.047	&	5.6	&	51	&	36 $\pm$ 12	\\
    & II	&	0.58	$\pm$	0.01 &	110	$\pm$	4	&	2.94	$\pm$	0.02	&	$<$	0.05	&	-0.054	$\pm$	0.015	&	3.6	&	14	&	38	$\pm$ 20\\

   & III	&	0.97	$\pm$	0.06 &	28.6	$\pm$	2.5	&	1.02	$\pm$	0.02	&	0.05	$\pm$	0.03	&	0.203	$\pm$	0.06	&	3	&	30	&	40 $\pm$ 18	\\
    & III	&	0.97	$\pm$	0.06 &	28.6	$\pm$	2.5	&	3.01	$\pm$	0.03	& $<$	0.06	&	-0.040	$\pm$	0.015	&	2.8	&	7	&	28 $\pm$ 15	\\

    & IV	&	2.5	$\pm$	0.3 &	8	$\pm$	2	&	1.05	$\pm$	0.03	&	0.06	$\pm$	0.02	&	0.115	$\pm$	0.038	&	2.9	&	14	&	33	$\pm$ 16\\
    & V	    &	2.7	$\pm$	0.3 &	7	$\pm$	2	&	1.05	$\pm$	0.02	&	0.06	$\pm$	0.02	&	0.127	$\pm$	0.05	&	2.8	&	20	&	37 $\pm$ 20	\\

9   & I	 &	0.53	$\pm$	0.01 &	148	$\pm$	5	&	2.89	$\pm$	0.02	&	$<$0.05	&	-0.062	$\pm$	0.02	&	3	&	11	&	36 $\pm$ 13	\\

   & I	&	0.53	$\pm$	0.01  &	148	$\pm$	5	&	3.72	$\pm$	0.04	&	0.20$\pm$0.08	&	-0.186	$\pm$	0.032	&	5.8	&	65	&	252	$\pm$ 55\\

10    & I	&	0.52	$\pm$	0.01  &	168.7	$\pm$	5.4	&	1.0	$\pm$	0.02	&	$<$0.08	&	0.30	$\pm$	0.11	&	2.7	&	29	&	28	$\pm$ 12 \\
    & I	&	0.52	$\pm$	0.01  &	168.7	$\pm$	5.4	&	2.95	$\pm$	0.03	&	0.12$\pm$0.04	&	0.191	$\pm$	0.053	&	3.6	&	50	&	127	$\pm$ 46 \\
      & II	&	0.67	$\pm$	0.02  &	71.3	$\pm$	3.7	&	1.02	$\pm$	0.01	&	$<$0.05	&	0.207	$\pm$	0.059	&	3.5	&	30	&	29 $\pm$ 36	\\
																							
11  & I	&	0.63	$\pm$	0.02  &	111.2	$\pm$	5	&	1.02	$\pm$	0.02	&	0.09$\pm$0.03	&	0.504	$\pm$	0.125	&	4	&	51	&	50 $\pm$ 12	\\
			
    & I	&	0.63	$\pm$	0.02 &	111.2	$\pm$	5	&	2.98	$\pm$	0.02	&	0.08$\pm$0.03	&	-0.182	$\pm$	0.048	&	3.8	&	50	&	95 $\pm$ 24	\\
    
    & I	&	0.63	$\pm$	0.02 &	111.2	$\pm$	5	&	3.65	$\pm$	0.06	&	0.35$\pm$0.08	&	-0.424	$\pm$	0.117	&	3.6	&	128	&	380 $\pm$ 84	\\

12  & I	&	0.62	$\pm$	0.01 &	106	$\pm$	3.5	&	1.04	$\pm$	0.01	&	$<$0.04	&	0.199	$\pm$	0.048	&	4.2	&	20	&	21 $\pm$ 7	\\

13  & I	&	0.58	$\pm$	0.01 &	146.2	$\pm$	4.2	&	2.90	$\pm$	0.02	&	$<$0.03	&	-0.066	$\pm$	0.02	&	3.3	&	12	&	35 $\pm$ 12	\\
																							
14  & I	&	0.55	$\pm$	0.01 &	138	$\pm$	4.7	&	1.02	$\pm$	0.01	&	0.04$\pm$0.02	&	0.283	$\pm$	0.070	&	4	&	33	&	30 $\pm$ 7	\\
    & I	&	0.55	$\pm$	0.01 &	138	$\pm$	4.7	&	2.93	$\pm$	0.03	&	0.08$\pm$0.03	&	-0.144	$\pm$	0.038	&	3.8	&	14	&	81 $\pm$ 22	\\
																				
    & I	&	0.55	$\pm$	0.01 &	138	$\pm$	4.7	&	3.81	$\pm$	0.02	&	$<$0.08	&	-0.06	$\pm$	0.02	&	3	&	22	&	79 $\pm$ 27 \\

     & II	&	0.84	$\pm$	0.02 &	52.3	$\pm$	2.3	&	0.99	$\pm$	0.01	&	$<$0.03	&	0.123	$\pm$	0.037	&	3.4	&	13	&	19	$\pm$ 8\\
																				
    & II	&	0.84	$\pm$	0.02 &	52.3	$\pm$	2.3	&	1.61	$\pm$	0.02	&	$<$0.03	&	-0.078	$\pm$	0.021	&	3.6	&	14	&	18 $\pm$ 8	\\
    
15  & I	&	0.52	$\pm$	0.01 &	175	$\pm$	5	&	1.07	$\pm$	0.02	&	$<$0.3	&	0.125	$\pm$	0.05	&	2.5	&	8	&	26 $\pm$ 12	\\
   & I	&	0.52	$\pm$	0.01 &	175	$\pm$	5	&	2.94	$\pm$	0.03	&	$<$0.1	&	-0.073	$\pm$	0.029	&	2.5	&	12	&	39 $\pm$ 27	\\

\hline
Pair-1	&	&	0.69	$\pm$	0.02 &	71.5	$\pm$	3	&	1.0	$\pm$	0.01	&	$<$0.03	&	0.148	$\pm$	0.035	&	4.2	&	21	&	15 	$\pm$ 5 \\
	&	&	0.69	$\pm$	0.02 &	71.5	$\pm$	3	&	1.6	$\pm$	0.01	&	$<$0.05	&	-0.096	$\pm$	0.022	&	4.4	&	27	&	19	$\pm$ 6 \\
	&	&	0.69	$\pm$	0.02 &	71.5	$\pm$	3	&	2.93	$\pm$	0.04	&	0.16$\pm$0.04	&	-0.195	$\pm$	0.037	&	5.2	&	47	&	94 $\pm$ 22	\\
Pair-2	&	&	0.62	$\pm$	0.02 &	95	$\pm$	4	&	1.05	$\pm$	0.01	&	0.03$\pm$0.01	&	0.25	$\pm$	0.04	&	6.3	&	61	&	28 $\pm$ 6	\\
	&	&	0.62	$\pm$	0.02 &	95	$\pm$	4	&	1.68	$\pm$	0.02	&	$<$0.06	&	-0.092	$\pm$	0.027	&	3.4	&	22	&	17 $\pm$ 5	\\
	&	&	0.62	$\pm$	0.02 &	95	$\pm$	4	&	2.96	$\pm$	0.01	&	$<$0.03	&	-0.105	$\pm$	0.0145	&	7.2	&	62	&	53 $\pm$ 10	\\
 	&	&	0.62	$\pm$	0.02 &	95	$\pm$	4	&	4.0	    $\pm$	0.04    &	0.04$^{+0.05}_{-0.03}$	&	-0.057	$\pm$	0.0183	&	3.1	&	26	&	62 $\pm$ 19	\\
Pair-3	&	&	0.56	$\pm$	0.01 &	148	$\pm$	7	&	1.01	$\pm$	0.01	&	0.04$\pm$0.02	&	0.219	$\pm$	0.063	&	3.5	&	33	&	20 $\pm$ 8	\\
	&	&	0.56	$\pm$	0.01 &	148	$\pm$	7	&	2.95	$\pm$	0.02	&	0.07$\pm$0.02	&	-0.117	$\pm$	0.026	&	4.5	&	39	&	59 $\pm$ 13\\
	&	&	0.56	$\pm$	0.01 &	148	$\pm$	7	&	3.70	$\pm$	0.05	&	0.2$\pm$0.09	&	-0.153	$\pm$	0.028	&	5.6	&	56	&	168	$\pm$ 41 \\
Pair-4	&	&	0.54	$\pm$	0.01 &	158	$\pm$	5	&	1.04	$\pm$	0.01	&	0.03$\pm$0.02	&	0.20	$\pm$	0.05	&	4	&	36	&	20	$\pm$ 6 \\
	&	&	0.54	$\pm$	0.01 &	158	$\pm$	5	&	2.93	$\pm$	0.02	&	0.09$\pm$0.02	&	-0.146	$\pm$	0.031	&	4.7	&	53	&	75 $\pm$ 16	\\
        &	&	0.54	$\pm$	0.01 &	158	$\pm$	5	&	3.68	$\pm$	0.05	&	0.20$\pm$0.09	&	-0.141	$\pm$	0.026	&	5.4	&	46	&	153	$\pm$ 43 \\ 

Pair-5	&	&	0.57	$\pm$	0.01 &	137	$\pm$	4	&	2.92	$\pm$	0.02	&	0.05$\pm$0.02	&	-0.085	$\pm$	0.021	&	4	&	28	&	49 $\pm$ 14	\\
        &	&	0.57	$\pm$	0.01 &	137	$\pm$	4	&	3.73	$\pm$	0.03	&	0.2$\pm$0.1	&	-0.192	$\pm$	0.022	&	8.7	&	130	&	252	$\pm$ 45 \\     
\hline
\end{tabular} 
\label{tab:burst-line-para}\\
\end{table*}

\subsection{Burst spectral features}

We searched for line features in five segments of all X-ray bursts. Evidence of narrow spectral residuals around 1~keV (emission), 1.7~keV (absorption), 3.0~keV (absorption), and 3.75 keV (absorption) is found in certain segments of the bursts. These residuals are fitted with Gaussian functions at their respective energies. We estimated the significance ($\sigma$) of the observed line features, defined as the ratio between the Gaussian normalization and its error at a 68\% confidence level. Furthermore, we also assessed the significance of these Gaussian line components using the {\it simftest} tool in {\tt XSPEC}. We generated 500 simulated spectra for each line component and analyzed the distribution of a change in the $\chi^2$ value ($\Delta\chi^2$) without and with the line feature in the model. We considered those burst segments where the $\Delta\chi^2$ value from the real data was larger than 95\% of the simulated realizations, indicating a p-value of less than 0.05. 
Table~\ref{tab:burst-line-para} presents the centroid line energy, width, and normalization parameters for the line features observed at a significance of $\ge$3$\sigma$ from the individual X-ray bursts.  Moreover,  the line parameters with significance between 2 and 3$\sigma$ are also presented for completeness. 
Additional parameters such as the blackbody temperature, blackbody emission radius (in km), change in the $\chi^2$ value ($\Delta\chi^2$) without and with the Gaussian line component, significance, and the line equivalent width (in eV) are also presented in the table. The observed blackbody radii are not color-corrected (see, e.g., \citealt{2013MNRAS.429.3266G, Suleimanov2017MNRAS.472.3905S}).

\begin{figure*}[ht!]
\centering
\includegraphics[height=4.7in, width=5.3in, angle=0]{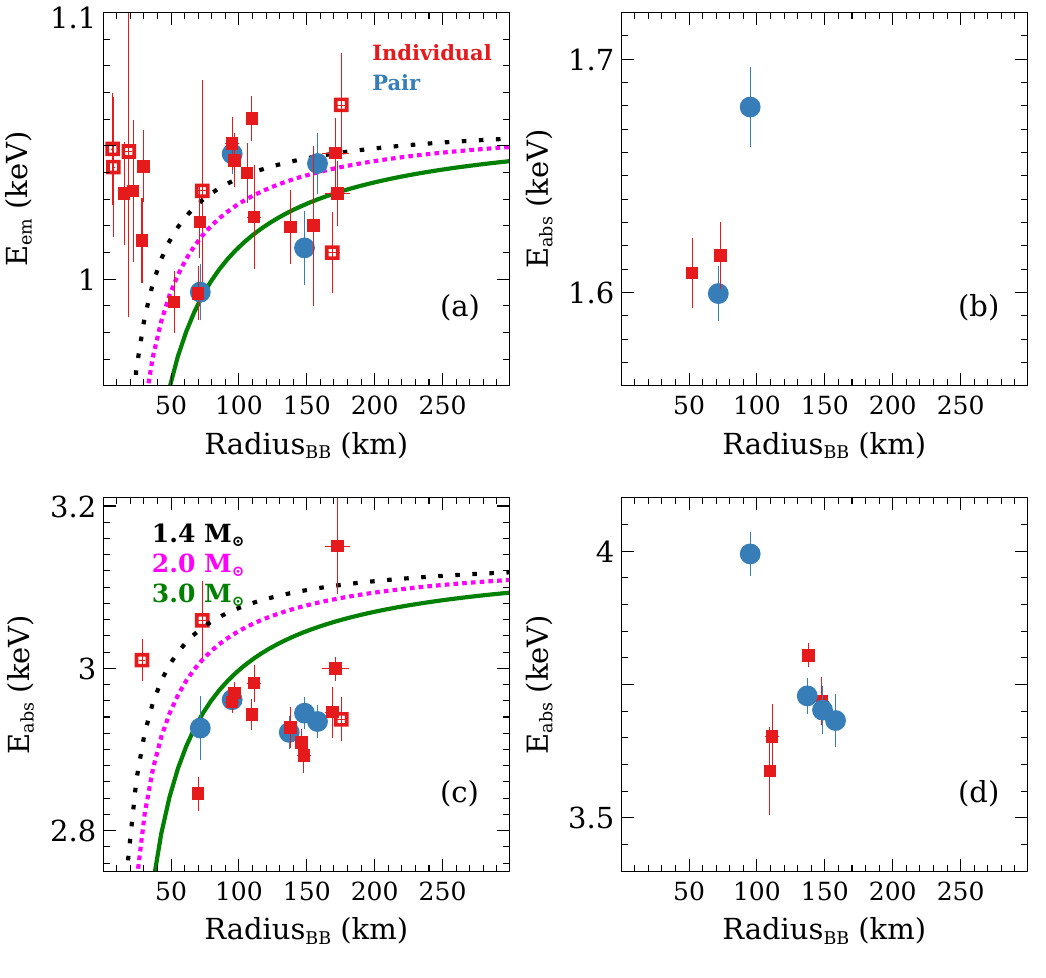}
\caption{The evolution of the emission line with energy around 1~keV, and three absorption lines with energies around 1.7, 3.0, and 3.75~keV, observed during X-ray bursts from \source is shown at different blackbody radii in panels (a), (b), (c), and (d), respectively.  The filled red squares represent values at a significance of $\ge$3$\sigma$ from individual bursts, while the hollow red squares show values below 3$\sigma$ level. The sky-blue circles correspond to the line energies obtained from the five pairs of burst spectra listed in Table~\ref{tab:burst-group}.  The black, magenta, and green curves represent the expected shift in the line energy due to the gravitational redshift for a NS mass of 1.4, 2, and 3\msol, respectively. The rest-frame energies for 1.0 and 3.0~keV features are assumed to be 1.06 and 3.14~keV, respectively.} 
\label{fig:Burst-line-para}
\end{figure*}

The evolution of line parameters with respect to the blackbody expansion radii of individual bursts is shown with red squares in Figure~\ref{fig:Burst-line-para}.  The 1.0~keV emission line, observed during the PRE and decay phases of most bursts, exhibits consistent energies ranging from 0.99 to 1.06~keV (panel (a) of Figure~\ref{fig:Burst-line-para}). Similarly, the absorption line energies near 3~keV remain almost constant within a narrow range between 2.84 and 2.98~keV, with an outlier at 3.14~keV  (panel (c) of Figure~\ref{fig:Burst-line-para}). The other two absorption lines, around 1.7 and 3.75~keV, were detected occasionally among 15 bursts. The absorption features around 1.7~keV ranged from 1.6 to 1.7~keV, while those near 3.75~keV varied between 3.6 and 3.70~keV, with an outlier around 4~keV, respectively  (panels (b) and (d) of Figure~\ref{fig:Burst-line-para}).  
In panels (a) and (c) of the figure, we also included the anticipated curves showing changes in the centroid line energy at different heights from the NS due to the gravitational redshift. The black, magenta, and green curves represent the expected changes in the line energy for a NS with 1.4, 2, and 3\msol, respectively. The rest-frame line energies for 1 and 3~keV lines were assumed to be 1.06 and 3.14~keV, respectively. These assumed values are the maximum energies of the respective features observed from these bursts (see Table~\ref{tab:burst-line-para}).   
%


\subsection{Co-added burst peak spectra} \label{co-add}

Following \citet{Strohmayer2019ApJ...878L..27S}, we also examined the peak spectrum (segment~I) of all 15 bursts without any Gaussian line parameters (see Table~\ref{tab:burst-group}). We identified five pairs of bursts with similar spectral parameters such as blackbody temperature, the expansion radius, and scaling factor \fa or photon index, and preferably paired those bursts that were close in time (see Table~\ref{tab:burst-group}). These pairs are named Pair-1 (Bursts~\#4 \& \#5), Pair-2 (Bursts~\#1 \& \#3), Pair-3 (Bursts~\#11 \& \#13), Pair-4 (Bursts~\#14, \& \#15), and Pair-5 (Bursts~\#6 \& \#9). This pairing approach enhances the significance of the line features observed around the PRE phases of the individual bursts. For reference, two of the pairs (Pair-1 and Pair-2) from the August 2017 observations were previously studied by \citet{Strohmayer2019ApJ...878L..27S}.

The combined peak spectra of each pair are then described using the ad-hoc power-law approach, consisting of a main blackbody and a power-law at fixed pre-burst emission parameters. The pre-burst emission parameters of these pairs were obtained by combining the respective pair persistent spectrum and fitting it with an absorbed Comptonization model with a blackbody component. Similar to individual bursts, we found the signature of line features in co-added burst spectra with enhanced line significance (see, Figure~\ref{fig:group-spec}).  For all five pairs of burst spectra, we also evaluated the significance of the Gaussian line components using {\it simftest}. The results indicated a higher level of significance, with a p-value of less than 0.001.  The line parameters for five pairs are shown in Table~\ref{tab:burst-line-para}.  The significance of line features was slightly increased in the paired spectra. The measurements from five pairs of co-added burst spectra at their peak (segment~I) are also shown in Figure~\ref{fig:Burst-line-para} with sky-blue circles.

To further enhance the significance and demonstrate that these features are unequivocally present around the PRE phase, we also co-added the peak spectra from bursts corresponding to five pairs, along with Burst~\#12. These eleven bursts exhibit expansion radii ranging from 74 to 170~km, with blackbody temperatures between 0.5 and 0.65~keV, based on the power-law modeling approach (see Table~\ref{tab:burst-group}). By fitting the combined spectra of all these eleven bursts (bottom panel of Figure~\ref{fig:group-spec}), we obtained the energies of emission and absorption line features to be at 1.02$\pm$0.01, 1.61$\pm$0.01, 2.94$\pm$0.01, and 3.74$\pm$0.03~keV with an equivalent width of 25$\pm$3, 32$\pm$4, 76$\pm$6, and 220$\pm$24~eV, respectively. The significance of these line features is estimated to be more than 10$\sigma$, based on the ratio of the line normalizations and their corresponding errors at a 68\% confidence level. We also found an absorption-like feature at 2.25$\pm$0.02~keV, with an equivalent width of 16$\pm$3~eV, and a significance of a 4.4$\sigma$.


\begin{figure}[]
\centering
\includegraphics[height=2.7in, width=3.4in, angle=0]{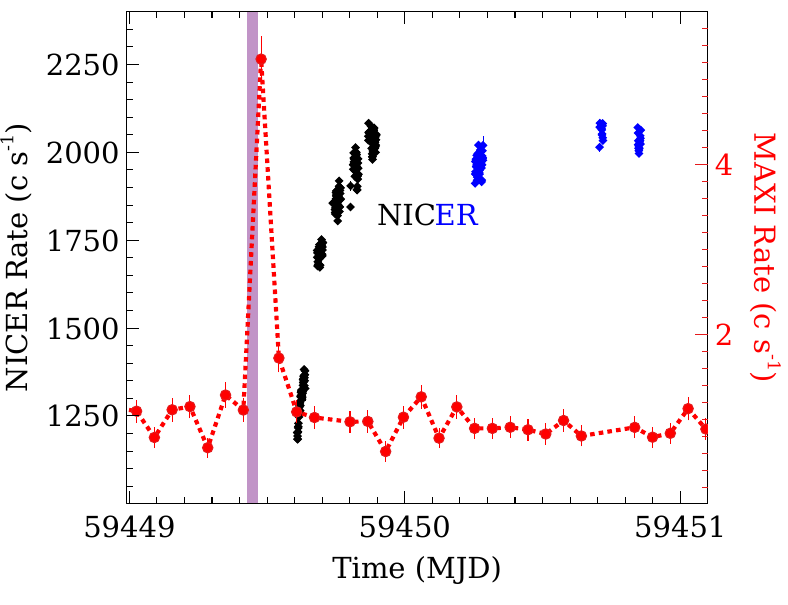} \\
\caption{The MAXI/GSC (red, 2--20 keV) and \nicer (0.5--10 keV;  black and blue) light curves around the long burst in 2021 August. MAXI detected the declining part of the long burst, while \nicer observed the emission from an aftermath phase where the count rate was recovering within half a day. \nicer data from ObsIDs 4050300105 and 4050300106 are shown at a 50~s time resolution in black and blue, respectively. The purple-shaded region in the figure indicates the possible peak time of the long burst, which occurred 3.14--4.70~hours before the start of the NICER observation.}  
\label{fig:aftermath-maxi-nicer}
\end{figure}
\begin{figure}[]
\centering
\includegraphics[height=3in, width=3.4in, angle=0]{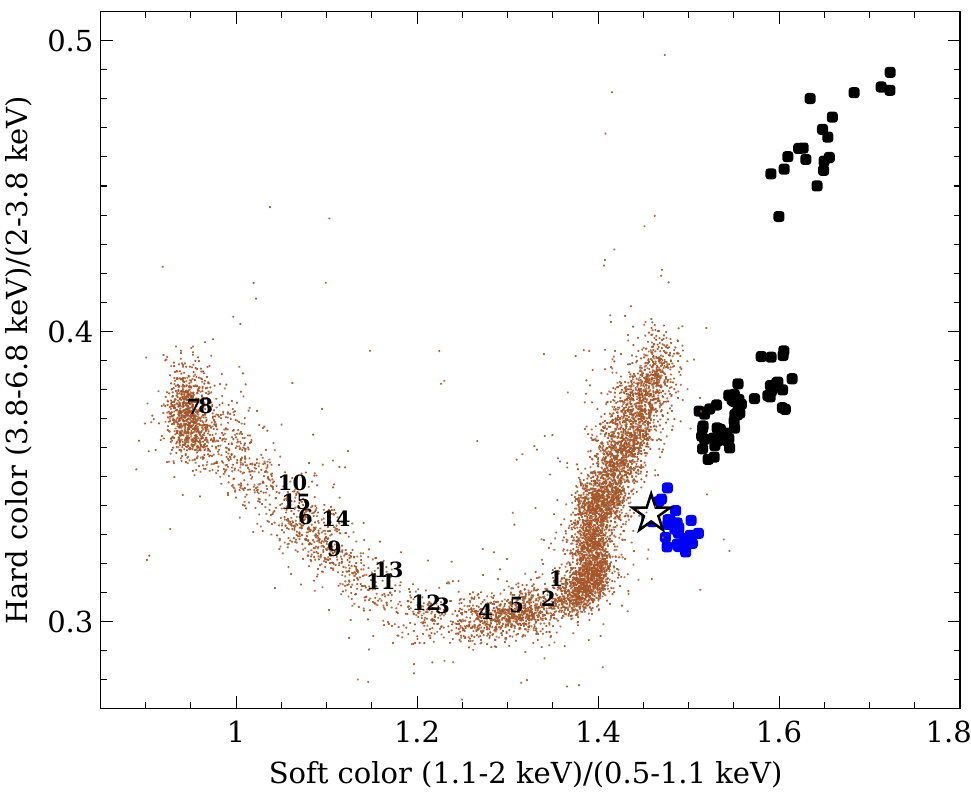}
\caption{The color-color diagram of \source from \nicer observations between  2017 June and 2023 August. The persistent emission before all 15 bursts is also marked as per their occurrence. The black and blue solid dots indicate the source state observed during the recovery phase after the long burst in 2021 August (see \nicer light curve in Figure~4). The star symbol corresponds to the last orbit data from this recovery phase. } 
\label{ccd}
\end{figure}

\begin{figure*}[]
\centering
\includegraphics[height=3.4in, width=7.2in, angle=0]{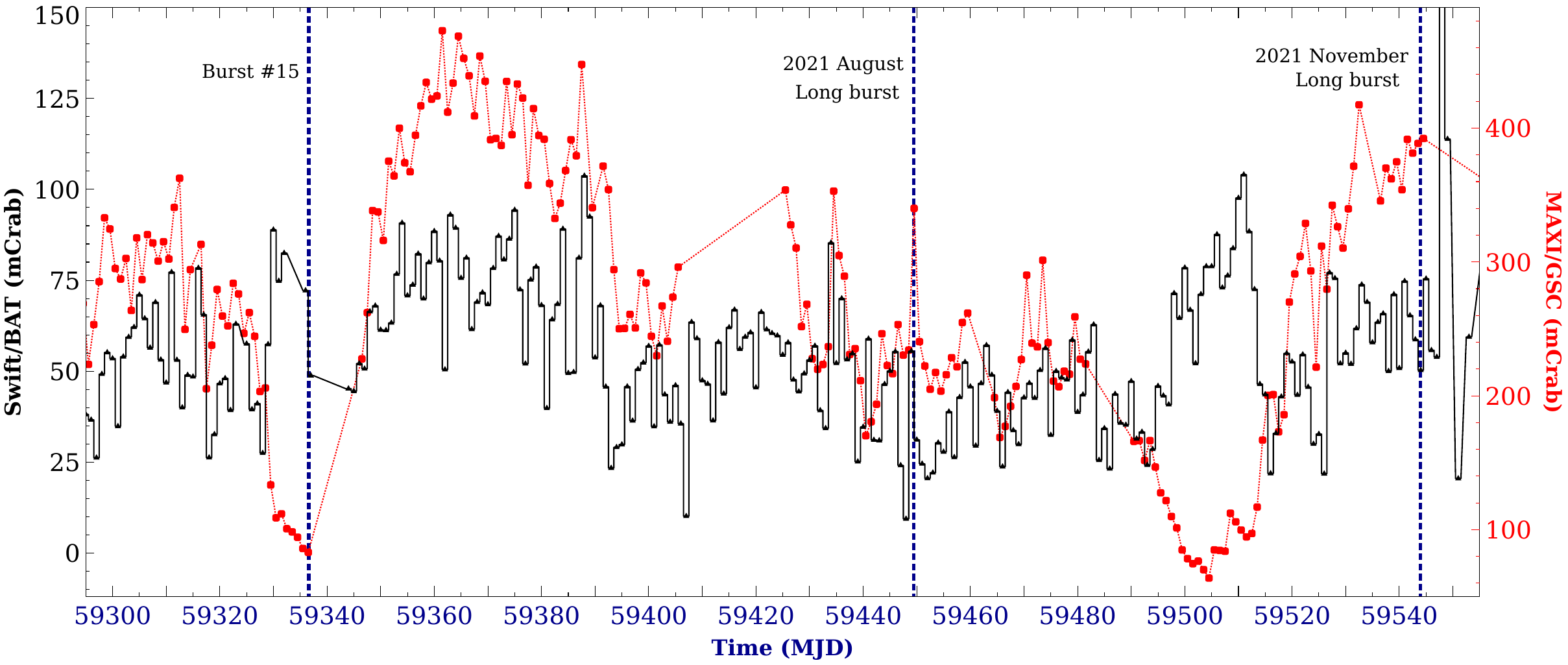}
\caption{MAXI/GSC (red square; 2-20 keV) and Swift/BAT (black; 15-50~keV) one-day averaged lightcurves of \source. The first dashed vertical line corresponds to the detection time of a thermonuclear X-ray burst (Burst~\#15) on 2021 May 02 in a low-hard state. The second and third vertical lines represent the time of long bursts observed by MAXI on 2021 August 23 and November 25, respectively. The low-hard states observed in 2021 April-May (around MJD 59336) and 2021 October (around MJD 59504) are separated by approximately 168~days. For the second long burst in 2021 November, see Appendix~\ref{discuss-superbursteffect} and \citet{Serino2021ATel15071....1S}. For reference, 1 Crab in MAXI/GSC and Swift/BAT corresponds to 3.8 and 0.220~c~s$^{-1}$~cm$^{-2}$, respectively.}  
\label{maxi-bat-lc}
\end{figure*}
\begin{figure}[]
\centering
\includegraphics[height=2.9in, width=3.3in, angle=0]{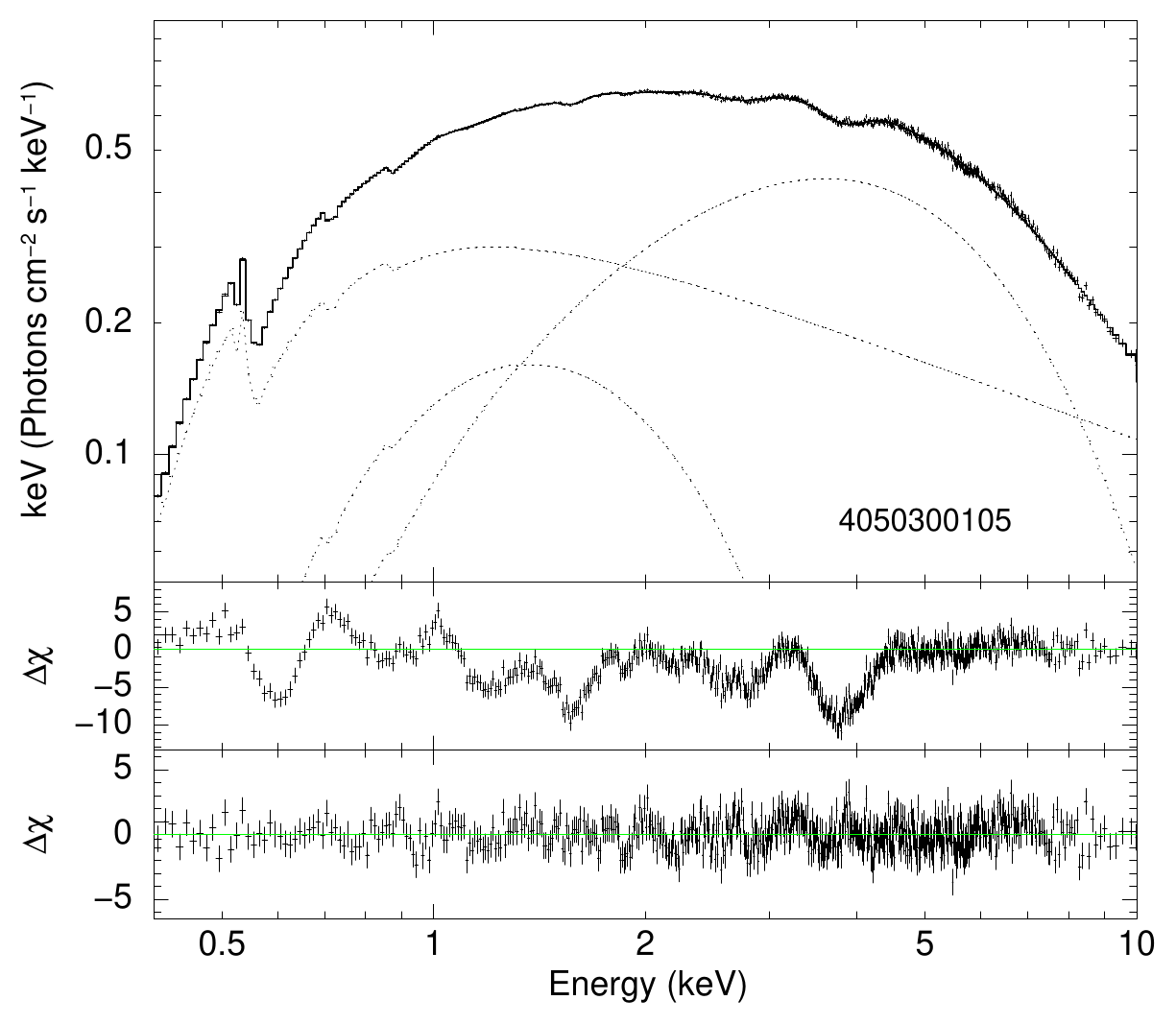} 
\includegraphics[height=2.9in, width=3.3in, angle=0]{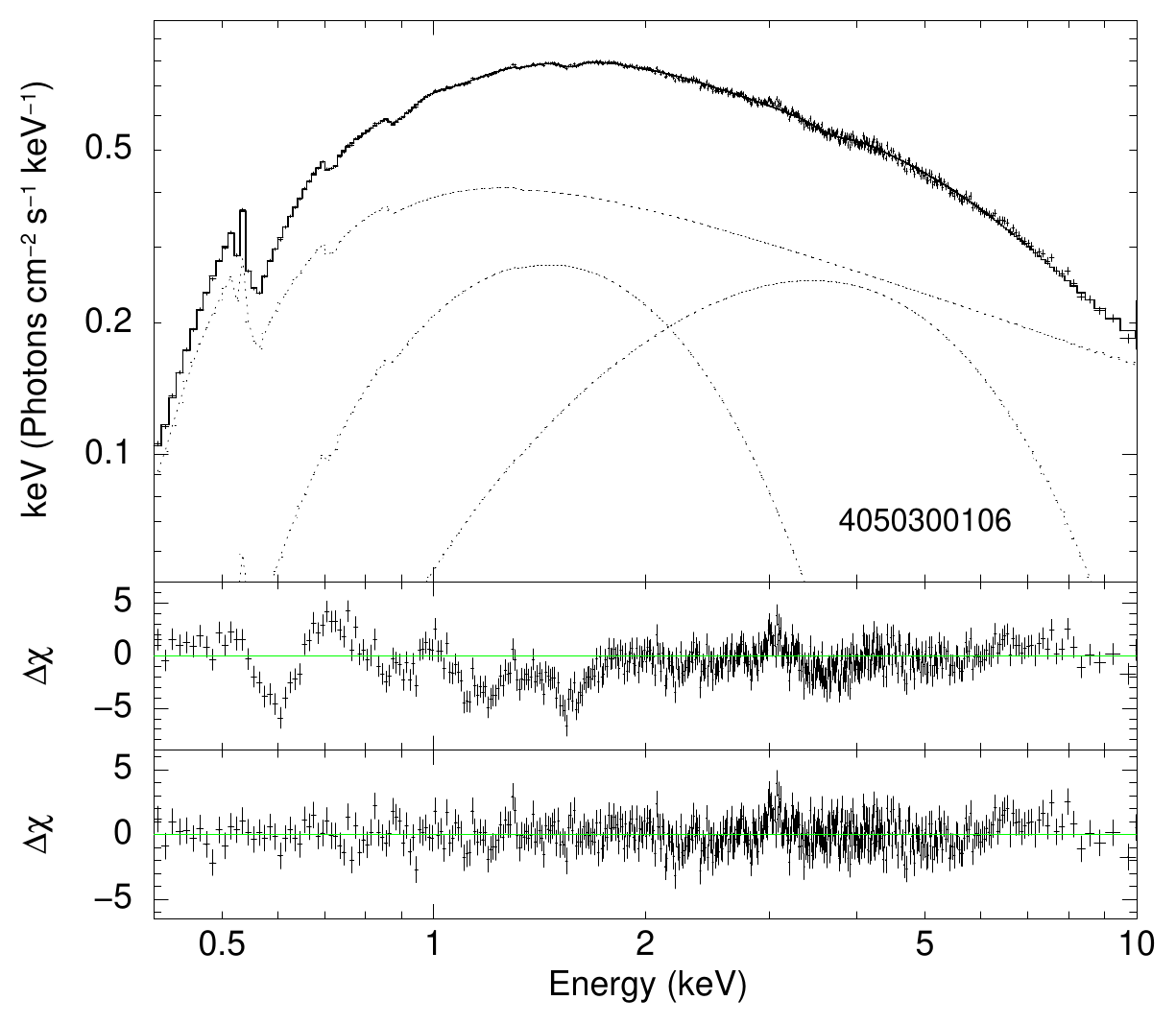} \\
\caption{The aftermath spectra of the 2021 August long burst from \source (ObsIDs 4050300105 and 4050300106) are fitted with an absorbed power-law model, two blackbody components, and multiple line features. Emission lines at 0.7 and 1.0 keV, absorption lines near 1.6, 2.7, and 3.8 keV, and an absorption edge at 0.53 keV are detected. The best-fit spectra and their corresponding residuals are presented in the top and bottom panels for each ObsID. The residuals, shown in the middle panels, are obtained by setting the normalization of each line and edge feature to zero.}
\label{fig:spec-after}
\end{figure}

\section{\nicer observations of a long X-ray burst aftermath emission} 
\label{spec-sec}

In addition to the narrow spectral features detected during X-ray bursts from \source, we present evidence of emission and absorption line features around similar energy ranges observed in the aftermath of a long burst in 2021 August. These findings are crucial for exploring potential connections between the line features detected during the short X-ray bursts and those observed after the long burst. They may help determine whether the same elements, such as those found in burst nuclear ashes, also dominate the composition of the accreted material.

\subsection{Flux recovery during the long burst aftermath } 

MAXI/GSC detected the cooling tail of a long X-ray burst from \source on 2021 August 23 (MJD~59449.477; \citealt{Serino2021ATel14871....1S}). 
The 2--20 keV MAXI light curve is shown in Figure~\ref{fig:aftermath-maxi-nicer}. Starting 3.14~hours after the MAXI trigger, \nicer monitored the source on 2021 August 23 and 24 with a total exposure of 13.5~ks. These data are stored under ObsIDs 4050300105 and 4050300106. The \nicer light curve in the 0.5--10~keV range increased from 1200 to 2000~c~s$^{-1}$ within 7.2 hours, between MJD 59449.60 and 59449.90 (black points in Figure~\ref{fig:aftermath-maxi-nicer}). A gradual stabilization of the count rate between 1900 and 2050~c~s$^{-1}$ was observed in the remaining three orbits of the \nicer data, covering 14.4 hours, between MJD 59450.26 and 59450.86 (blue points in Figure~\ref{fig:aftermath-maxi-nicer}). This provides a total observational coverage of about 1.26 days with NICER. Such post-burst recovery after an intermediate duration burst has been seen as well in IGR~J17062-6143, but on a longer timescale \citep{Bult2021ApJ...920...59B}. 

To examine the evolution of the source state during the aftermath emission, a color-color diagram was created using \nicer data accumulated between June 2017 and August 2023 (Figure~\ref{ccd}). The soft color is defined as the ratio of count rates in the 1.1--2.0 and 0.5--1.1~keV energy bands. We estimated the hard color as the ratio of count rates between 3.8--6.8 and 2.0--3.8~keV energy bands  (e.g. \citealt{Bult2018, Jaisawal2024}). We also included the colors before all 15 bursts, which identifies the low-hard bursting state of the source. 
The black and blue points correspond to the \nicer aftermath emission on 2021 August 23 and 24, respectively.  The aftermath emission appears distinct from the usual accretion state of the source (brown). The soft and hard colors during the aftermath emission evolved from 1.72 and 0.48 to 1.46 and 0.33, respectively. The star symbol in the color-color diagram represents data from the last orbit of NICER observations (ObsID 4050300106), suggesting the aftermath emission was still off the main track at the end of the observation. 

We further analyzed the long-term MAXI/GSC \citep{Mihara2011PASJ...63S.623M} and Swift/BAT \citep{Krimm2013ApJS..209...14K} light curves of \source to evaluate the accretion state before the long burst (Figure~\ref{maxi-bat-lc}). Consistent with a high-soft or intermediate accretion state, the source intensities on MJD 59440--59460 were found to be $>$200 and $<$60~mCrab in the 2--20 and 15--50 keV energy ranges, respectively. Hence, \source was clearly outside the low-hard bursting state before the long burst in 2021 August. Unlike usual type-I bursts, the event observed by MAXI may be related to a superburst whose occurrence does not depend on the source bursting state (see  \citealt{Strohmayer2002ApJ...566.1045S, Zand2012A&A...547A..47I} for the discussion of the superburst observed from \source with \rxte on 1999 September 9). Additional supporting evidence is the last X-ray burst (Burst~\#15) observed with \nicer on 2021 May 2 (MJD~59336.608), which occurred in a low-hard state almost 113 days before the long burst. Low-hard states of \source are typically observed at intervals of about 170--176 days \citep{Chou2001ApJ...563..934C, Strohmayer2002ApJ...566.1045S}. Figure~\ref{maxi-bat-lc} shows the presence of low-hard states in 2021 April-May and 2021 October with a time interval of 168~days. Thus, the long burst in 2021 August was out of phase with respect to the usual bursting state of \source, and was likely a superburst. See more discussion in Appendix~\ref{discuss-superbursteffect} on this superburst and the quenching effect.

\begin{table}[]
\centering
\scriptsize
\caption{Spectral parameters from the aftermath emission of the long burst in 2021 August, obtained by fitting two different models: an absorbed Comptt with a blackbody component (Model-I), and two blackbody components with a power-law (Model-II).  Two emission lines, three absorption lines, and an absorption edge are detected in the spectra. The errors are quoted at a 68\% confidence range.  
}
\begin{tabular}{ccc  cc}
\hline
 &  \multicolumn{2}{c|}{4050300105} & \multicolumn{2}{c}{{4050300106}}  \\
\hline
Parameters              &Model-I        &Model-II        &Model-I   &Model-II \\
\hline
N$\rm{_H}$ (10$^{21}$ cm$^{-2}$)         &2.3$\pm$0.1     &2.0$\pm$0.1   &2.1$\pm$0.1     &2.0$\pm$0.1    \\
Comp$\rm_{T0}$ (keV)           &0.05$\pm$0.02  &--  &0.04$\pm$0.01   &--     \\
Comp$\rm_{kT}$ (keV)           &2$\pm$0.1  &-- &2.1$\pm$0.1 &--     \\
$\tau$ (keV)                 &10.2$\pm$0.1   &--  &9.8$\pm$0.2 &--\\
Norm$\rm_{Comp}$           &2.1$\pm$0.3    &--    &2.2$\pm$0.4  &--\\
kT$\rm _{BB1}$ (keV)   &1.0$\pm$0.01    &1.26$\pm$0.01   &0.62$\pm$0.01         &1.20$\pm$0.02       \\
R$\rm _{BB1}$ (km)     &10.7$\pm$0.2   &10.2$\pm$0.2    &25$\pm$1     &8.3$\pm$0.3\\
kT$\rm _{BB2}$ (keV)   &--   &0.42$\pm$0.01    &--   &0.45$\pm$0.01 \\
R$\rm _{BB2}$ (km)     &--  &36$\pm$3       &--  &41$\pm$4\\
$\Gamma$               &--  &1.59$\pm$0.03    &--  &1.55$\pm$0.03\\
Norm$_\Gamma$           &--  &0.42$\pm$0.02     &--  &0.56$\pm$0.02\\

\\
Emission lines: \\
Energy (keV)    &0.71$\pm$0.02      &0.67$\pm$0.02   &0.67$\pm$0.02     &0.67$\pm$0.02\\
Width (keV)     &0.02$\pm$0.01      &0.10$\pm$0.01   &0.08$\pm$0.02     &0.10$\pm$0.01\\
Norm  (10$^{-3}$) &3.6$\pm$0.8       &33.7$\pm$8     &22$\pm$7      &39.5$\pm$9\\
\\
Energy (keV)     &1.01$\pm$0.01   &0.98$\pm$0.01   &0.98$\pm$0.02   &0.98$\pm$0.01\\
Width (keV)     &0.03$\pm$0.01      &0.07$\pm$0.01    &0.06$\pm$0.02    &0.06$\pm$0.01\\
Norm (10$^{-3}$) &2.1$\pm$0.2    &6.3$\pm$0.8    &5.5$\pm$1.5     &5$\pm$1\\
\\
\\
Absorption \\
features: \\
Energy (keV)    &1.59$\pm$0.01    &1.58$\pm$0.01    &1.57$\pm$0.01    &1.56$\pm$0.01\\
Width (keV)     &0.05$\pm$0.01      &0.06$\pm$0.01   &0.07$\pm$0.03    &0.05$\pm$0.01\\
Norm (10$^{-3}$)  &-1.8$\pm$0.2      &-2.1$\pm$0.2   &-3$\pm$0.9   &-1.8$\pm$0.3\\ 
\\
Energy (keV)   &2.61$\pm$0.01    &2.70$\pm$0.01   &--   &--\\
Width (keV)    &0.42$\pm$0.02    &0.20$\pm$0.01 &--   &--\\
Norm (10$^{-3}$) &-21.2$\pm$2  &-4.31$\pm$0.42   &--   &--\\
\\
Energy (keV)    &3.78$\pm$0.01  &3.80$\pm$0.01   &3.57$\pm$0.04    &3.64$\pm$0.03\\
Width (keV)     &0.30$\pm$0.01  &0.26$\pm$0.01   &0.30$\pm$0.05     &0.20$\pm$0.02\\
Norm (10$^{-3}$)  &-13.6$\pm$0.4    &-9.2$\pm$0.42  &-7.4$\pm$2   &-2.2$\pm$0.4\\

\\
E$\rm_{edge}$ (keV)  &0.53$\pm$0.01 &0.56$\pm$0.01   &0.56$\pm$0.01    &0.56$\pm$0.01\\
$\tau\rm_{edge}$     &0.10$\pm$0.01 &0.17$\pm$0.02   &0.10$\pm$0.03   &0.16$\pm$0.02\\

\\
Flux (10$^{-9}$)     &7.39$\pm$0.01             &7.30$\pm$0.01    &7.40$\pm$0.02    &7.36$\pm$0.01\\
$\chi^2_\nu$ (dof)  &1.36 (935)          &1.23 (935)   &1.07 (938)    &1.16 (938)\\
\hline
\end{tabular}
\label{tab:aftermath}\\
\end{table}

\begin{figure}[]
\centering
\includegraphics[height=4.5in, width=3.35in, angle=0]{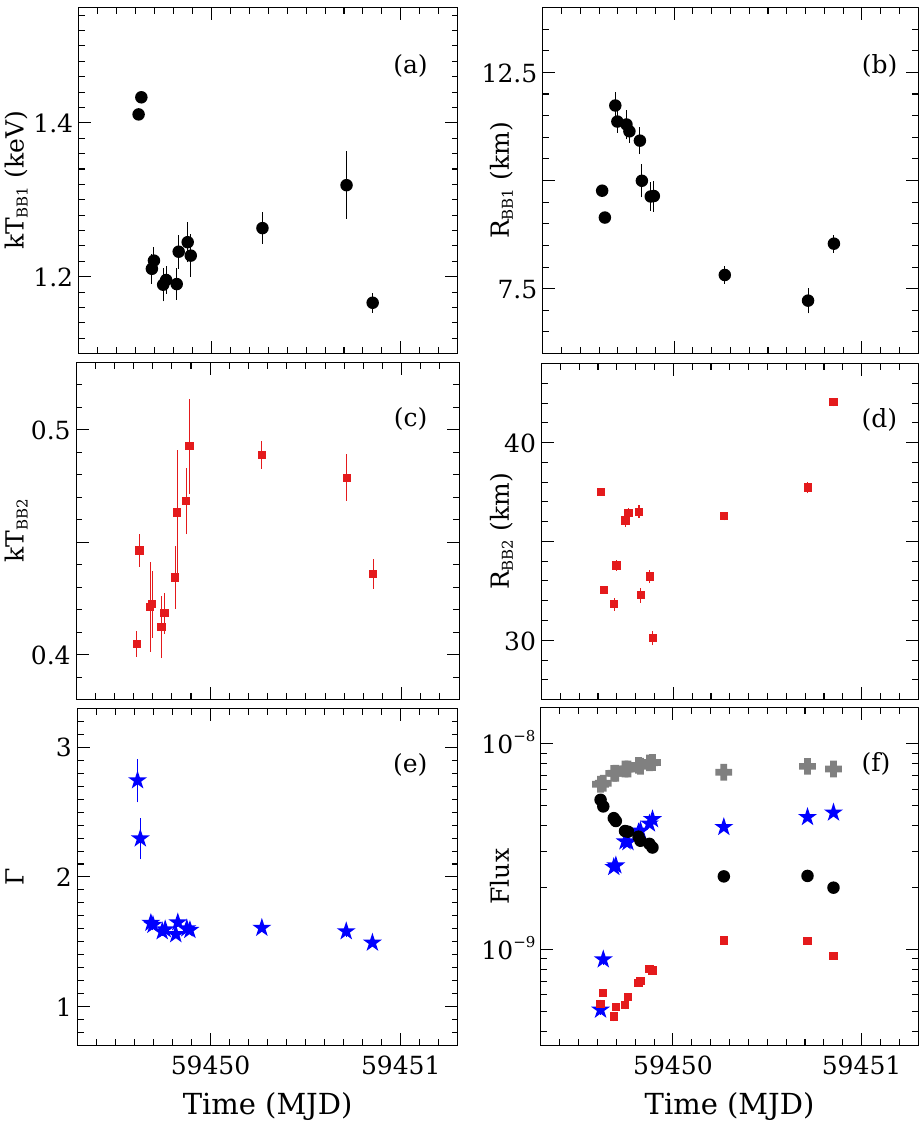} \\
\caption{Spectral evolution of the aftermath emission of 2021 August long burst fitted by an absorbed power-law model along with two blackbodies (Model-II) and multiple narrow line components. The obtained parameters are shown in panels (a)--(f). The 0.5--10 keV fluxes from individual components such as hotter blackbody (black), cooler blackbody (red), and power-law (blue) are also presented, along with unabsorbed total model flux (grey color, panel~(f)) in erg~s$^{-1}$~cm$^{-2}$ unit.} 
\label{fig:spec-par-aftermath2}
\end{figure}

\begin{figure}[]
\centering
\includegraphics[height=6.4in, width=3.2in, angle=0]{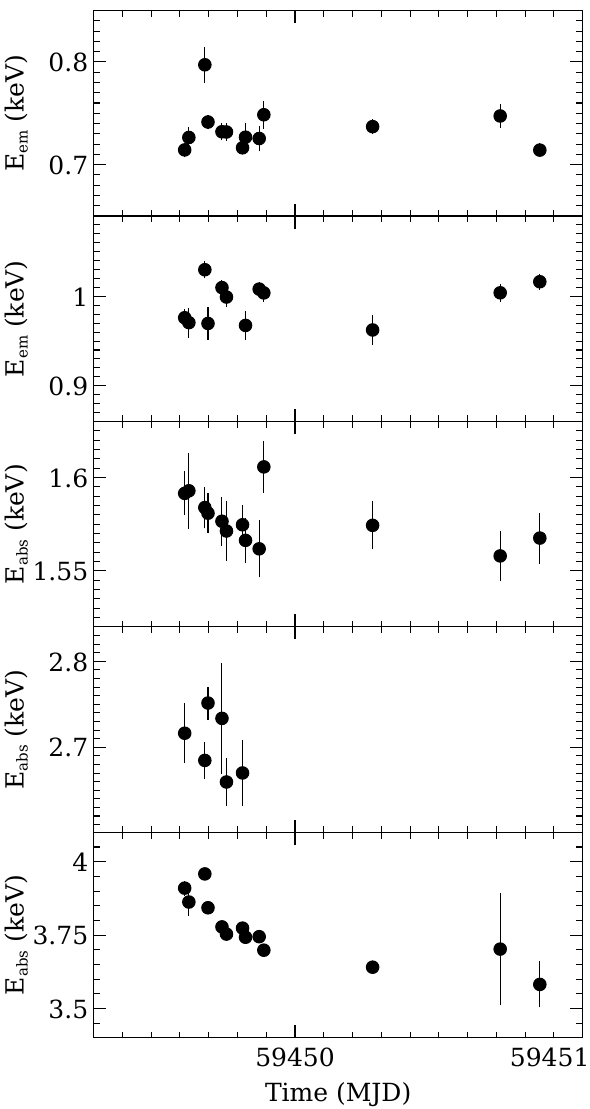} 
\caption{Evolution of emission lines around 0.7 and 1.0 keV, and absorption lines near 1.6, 2.7, and 3.8~keV during the aftermath emission of a long burst in 2021 August.} 
\label{fig:spec-bb-po-aftermath-line}
\end{figure}


\subsection{Detection of spectral features in the aftermath emission of the 2021 August long burst} \label{spec-sec-aftermath}

We next performed spectral studies in the 0.4--10 keV range to investigate the evolution of the aftermath emission following the long burst (candidate superburst) on 2021 August 23. We consider an absorbed Comptonization model with a blackbody component (Model-I) to fit the evolving persistent emission from ObsIDs 4050300105 and 4050300106 with exposure times of 9.6 and 3.9~ks, respectively.  Model-I is the same as the pre-burst persistent continuum model discussed in Section~3. The absorption column density is described by the {\tt Tbabs} component in {\tt XSPEC} with {\tt wilm} abundances \citep{Wilms2000} and {\tt verner} cross-sections \citep{Verner1996ApJ...465..487V}. However, the model failed to adequately fit the spectra due to multiple narrow spectral residuals below 4~keV. We found evidence of two emission lines at 0.7 and 1~keV, and three absorption lines at 1.6, 2.7, and 3.8~keV. The absorption feature around 3.8~keV is most prominently observed in the spectrum from ObsID 4050300105.   An Oxygen absorption edge at 0.53 keV was also detected in the spectra (see, e.g., \citealt{Rogantini2021A&A...645A..98R}). The inclusion of Gaussian components for the spectral features results in a fit with a reduced $\chi^2$ of 1.36 (Table~\ref{tab:aftermath}). 

In addition to the Comptonization-based model, we examined the aftermath spectra using a simple absorbed power-law model with a blackbody component. However, due to the spectral residuals below 2 keV, this model could not describe the continuum properly. To address this, an additional blackbody component was included, leading to an improved fit for both ObsIDs. This model also required two emission lines, three absorption lines, and an absorption edge to account for the observed features. Incorporating these components significantly improved the reduced chi-squared value (see Table~\ref{tab:aftermath}).
Figure~\ref{fig:spec-after} shows the best-fit energy spectrum (top) and corresponding residuals (bottom) for each ObsIDs, modeled with the absorbed power-law and two blackbody components (Model-II), along with an edge and line features. Moreover, the (respective) middle panel displays the residuals with the normalizations of the edge and line features set to zero, highlighting the strength of these spectral features relative to the best-fit continuum model.

To examine the detailed evolution of the aftermath emission and spectral features, we divided the data from ObsIDs 4050300105 and 4050300106 into 13 time-resolved intervals. The exposure time for each interval ranged between 500 to 2500~s, with a median around 900~s. The time-resolved spectra were fitted using the simpler model i.e. Model-II, to resolve the contributions from thermal and non-thermal components potentially originating from the NS, accretion disk, and coronal environment. These intervals also revealed evidence of two emission lines at 0.7 and 1~keV, and three absorption lines near 1.6, 2.7, and 3.8~keV. The addition of Gaussian line components provided an acceptable fit, with reduced chi-squared values between 0.93 and 1.16. These line features were detected with a significance $>3\sigma$, calculated as the ratio between the Gaussian normalization and its error at a 68\% confidence level.

From Model-II, one of the blackbody temperatures was initially high and then cooled down from 1.4 to 1.1~keV during the \nicer observations (panel~(a) of Figure~\ref{fig:spec-par-aftermath2}). The corresponding emission radius, shown in panel~(b) of the figure, varied between 7.5 and 12.5~km (not color-corrected; assuming a distance of 8.4~kpc), suggesting the emission is most likely from the NS surface which is still hot after the long burst. The flux from this hotter blackbody component declined from 5.3$\times$10$^{-9}$ to 2$\times$10$^{-9}$~erg~s$^{-1}$~cm$^{-2}$ during the observation in the 0.5--10 keV range (panel~(f) of Figure~\ref{fig:spec-par-aftermath2}). The second blackbody component was found to be relatively cooler with a temperature and emission radius, varying in a range of 0.4--0.5~keV and 30--40~km, respectively (panel~(c) and (d) of the figure). This component may represent the emission from the inner accretion disk. The photon index evolved from 3 to 1.5 during the observation (panel~(e)). A softer photon index is anticipated due to coronal cooling as an immediate interaction between the long burst and the accretion environment. The non-thermal power-law component may also come from the reflection of inner disk photons from parts of the outer disk. The power-law and soft blackbody fluxes increased during the \nicer observations (panel~(f) of Figure~\ref{fig:spec-par-aftermath2}), likely representing the gradual recovery of emissions from the corona and accretion disk. We observed a moderate increase in the second blackbody radii during the recovery phase. This may indicate that the strengthening of the corona and accretion boundary layer may have pushed the inner accretion disk farther from the surface. Furthermore, as the accretion disk restores during the recovery phase, increased emission from the outer regions of the inner disk suggests a progressive expansion of the disk radius.    
The column density was found in a range of (1.9--2.2)$\times10^{21}$~cm$^{-2}$ (not shown in the figure).

Figure~\ref{fig:spec-bb-po-aftermath-line} shows the evolution of emission and absorption lines, obtained from time-resolved spectroscopy of the aftermath emission using Model-II. The 0.7 and 1.0 keV emission lines remained nearly constant during the recovery phase, with energies in the ranges of (0.7--0.8) and (0.96--1.02)~keV, respectively. The centroid line energies of the absorption feature around 1.6, 2.8, and 4~keV decreased during this period, with line energies in the ranges of (1.55--1.61), (2.64--2.75), and (3.58--4.0), respectively. Furthermore, the observed 1~keV emission line and three absorption lines during aftermath emission also overlap with the energy ranges of detected features from the short X-ray bursts as shown in Figure~\ref{fig:Burst-line-para}.

\section{Discussion} \label{discuss}

We studied time-resolved spectra of 15 thermonuclear X-ray bursts from \source using \nicer, in search of spectral features that may originate from nuclear-burning ashes and/or the accretion environment. Additionally, we examined the aftermath emission of a long burst (candidate superburst) that occurred in 2021 August. Similarly to the X-ray bursts, the aftermath emission displays a 1~keV emission line and three absorption features. The implications of our findings are discussed in the following sections.

\subsection{Evidence of heavy nuclear ashes during X-ray bursts from \source}

Thermonuclear X-ray bursts are powerful explosions on the NS surface, driven by unstable nuclear burning of accreted material. For pure helium accretion as expected in 4U 1820--30, the nuclear burning is dominated by the triple-$\alpha$-reaction, $\alpha$-captures, and the $\alpha$p-process that is a sequence of ($\alpha$,p) and (p,$\gamma$) reactions catalyzed by trace amounts of hydrogen produced by the reactions. The resulting composition of the produced nuclei is mainly limited to $\alpha$-chain isotopes in the mass number range of $A=28-60$ \citep{Schatz2001PhRvL..86.3471S, Woosley2004ApJS..151...75W,  Weinberg2006ApJ...639.1018W, 2023A&A...678A.156H}.

The heavy nuclear ashes produced during X-ray bursts can settle at shallow depths through convective mixing \citep{Weinberg2006ApJ...639.1018W}, or be buried by freshly accreted material after the explosion. However, during PRE bursts, strong radiative winds can expose these ashes. These winds typically originate at shallow column depths but rapidly expel material from deeper surface layers at depths $\ge$ 10$^6$ g~cm$^{-2}$ \citep{Yu2018ApJ...863...53Y}. Depending on the ignition depth, the winds can last from a few seconds to over a hundred seconds, carrying out about 0.2\% of the total accreted material \citep{Yu2018ApJ...863...53Y}. Lighter elements like helium and carbon are expelled first from the upper surface layers, followed by heavier nuclear ashes (mass number $>$ 40) from deeper layers about a second later. Detection of such elements from X-ray bursts would provide valuable insights into the nuclear burning processes. This may also allow us to estimate the gravitational redshift, which depends on the mass and radius of the NS, leading to an estimation of its compactness. 

By studying two pairs of X-ray bursts around the PRE phases (Pair-1 and Pair-2 in Table~\ref{tab:burst-group}) from \source, \citet{Strohmayer2019ApJ...878L..27S} discovered a 1~keV emission line and two absorption lines at 1.7 and 3~keV. These lines were systematically red-shifted as the blackbody expansion radii decreased, suggesting either a gravitational redshift, a wind-induced blue-shift, or a combination of both effects. In our study, we report the detection of narrow spectral features around 1.0, 1.7, 3.0, and 3.75~keV from a larger sample of X-ray bursts from \source (Figure~\ref{fig:Burst-line-para}).

The composition of the ejected nuclear ashes is expected to be dominated by $\alpha$-elements such as $^{28}$Si, $^{32}$S, $^{36}$Ar, $^{40}$Ca, $^{44}$Ti, and $^{48}$Cr, with a mass fraction $\ge$10$^{-3}$  in the case of He-powered X-ray bursts \citep{Weinberg2006ApJ...639.1018W, Yu2018ApJ...863...53Y}. 
Among these elements, Si exhibits K$\alpha$ emission at 1.74, 1.87, and 2.00~keV (rest-frame), corresponding to its neutral, He-like (Si~XIII), and H-like (Si~XIV) states, respectively \citep{Hoof2018Galax...6...63V}. Similarly, neutral Ar, He-like (Ar XVII), and H-like (Ar XVIII) have K$\alpha$ emissions at approximately 2.96, 3.14, and 3.32~keV, respectively \citep{Hoof2018Galax...6...63V}. For Ca, the neutral, He-like (Ca XIX), and H-like states (Ca XX) have K$\alpha$ transitions near 3.69, 3.90, and 4.10~keV, respectively \citep{Hoof2018Galax...6...63V}.
Based on the absorption features detected during X-ray bursts of \source, we propose that the lines around 1.7, 3.0, and 3.75~keV may originate from the burst nuclear ashes of $^{28}$Si, $^{36}$Ar, and $^{40}$Ca, respectively.  
The K$\alpha$ line energy generally increases with ionization, from neutral to fully ionized states, with this effect being more pronounced for heavier elements (see, e.g., \citealt{Kallman2004ApJS..155..675K}). It is unlikely that these wind ashes are in a neutral state during the PRE phase (see also \citealt{Barra2025A&A...694A.266B}). In our study, the detected absorption energies are mostly below or near the respective neutral K$\alpha$ transitions, suggesting the possible influence of the NS's gravitational redshift.

In principle, the absorption lines may also be affected by wind-induced Doppler shifts. From time-resolved spectroscopy of X-ray bursts \citep{Jaisawal2024}, we can estimate the average velocity (v) of the NS photosphere to be $\approx$250~km~s$^{-1}$, based on changes in photospheric radii over a time duration. The highest photospheric velocity was found to be $\approx$2200~km~s$^{-1}$ during Burst~\#8, where the NS photosphere expanded to almost 1000~km, using fine time-resolved spectroscopy with the variable persistent emission method. This photospheric velocity is consistent with the super-Eddington wind velocity that is usually below 0.01~c \citep{Yu2018ApJ...863...53Y}. Therefore, the anticipated change in line energy due to the Doppler shift can be estimated as $\Delta$E/E = v/c, which gives 0.007 for a velocity of 2200~km~s$^{-1}$. This corresponds to a change in line energies of less than 7 and 30~eV for 1 and 4~keV line features, respectively. Given \nicer's spectral resolution of 85~eV at 1~keV, the Doppler shift due to photospheric velocity becomes negligible, but measurable considering the centroid line significance. Thus, a combination of wind-induced Doppler (blue)shift and the effect of NS gravitational redshift can explain the evolution of absorption line features from the ejected ashes during the PRE bursts. 

\subsection{Absorption lines from the aftermath emission: metal-rich accretion flow?}

We report the first detection of absorption lines at energies around 1.61, 2.75, and 4.0~keV from \source during the accretion-driven aftermath emission of a candidate superburst. This finding raises two key questions: 1) Do the features originate from the accreted material? 2) Why was a prominent absorption line detected around 4~keV instead of the 6--7~keV line sometimes seen in the persistent emission? The companion in \source is a He white-dwarf based on the short orbital period (11.4 minutes; \citealt{Stella1987ApJ...312L..17S, Rappaport1987ApJ...322..842R}), and is expected to provide He-rich, metal-deficient material to the accretion flow. This metal deficiency can be supported by the high surface gravity (log~g$\sim$8) of white dwarfs, which causes heavier species to sink into its interior layers, although convective mixing may still occur (see, e.g., \citealt{Kepler2019MNRAS.486.2169K}). Studies of ultra-compact X-ray binary systems using X-ray, UV, and optical spectroscopy have identified potential element signatures from He, C, N, and O, but no elements likely heavier than Ne \citep{Nelemans2010MNRAS.401.1347N}. These findings suggest that the accreted material predominantly consists of lighter elements from the He white-dwarf companion, with relatively lower abundances of heavier species. Thus, the observed absorption features, particularly the strong 4.0~keV line, are unlikely to originate from the metal-poor accretion flow from the companion star in \source system.

The 2021 August long burst is a candidate superburst, occurring more than two decades after the first well studied superburst from \source discovered in 1999 with \rxte \citep{Strohmayer2002ApJ...566.1045S}. This event was a remarkable three-hour phenomenon, showing super-expansion during the PRE phases of both the precursor and the superburst \citep{Keek2012ApJ...756..130K}. Superbursts typically exhibit a precursor, a short X-ray burst powered by explosive helium burning. The precursor is triggered by the shock launched by the superburst ignition in deeper carbon-rich layers. It is expected to be a strong PRE burst \citep{Keek2012ApJ...756..130K}, producing elemental abundances similar to those seen in type-I bursts. 

A PRE superburst may eject significant amounts of nuclear ashes, either from the precursor or from elements synthesized by the superburst itself \citep{Weinberg2006ApJ...639.1018W}.
During superbursts, the high temperatures in deeper burning layers convert the composition into iron-group nuclei, approaching nuclear statistical equilibrium, while shallower regions experience moderate temperatures and undergo nucleosynthesis via explosive helium burning. Helium is produced via the $^{12}$C($^{12}$C,$^4$He)$^{20}$Ne reaction \citep{2011ApJ...743..189K} and then re-captured. The model studies indicate that the main products are dominated by $\alpha$-elements up to $^{32}$S, though these calculations do not account for a PRE superburst \citep{2011ApJ...743..189K}. It is possible that more energetic superbursts could produce heavier $\alpha$-species, resulting in nuclear ashes similar to those from the regular short bursts.

We propose that the wind mass loss during the 2021 superburst and/or its precursor enriched the accretion flow with nuclear ashes, enhancing its metallicity. As the accretion disk and corona emission recovered, this metal-rich flow from the accretion atmosphere or inner disk wind became exposed to the hot NS surface, producing the observed element absorption features during the recovery phase. This provides potential evidence for metal enrichment of the accretion flow due to burst wind loss. A decrease in line energies during the aftermath phase may suggest the effects of the NS gravitational field (redshift) on the line-forming regions. 

Assuming an ignition depth of (0.1--1)$\times$10$^{12}$~g~cm$^{-2}$ for the superburst \citep{Strohmayer2002ApJ...566.1045S, Keek2012ApJ...756..130K, Zand2017symm.conf..121I}, we estimate the wind mass loss rate to be (6--10)$\times$10$^{18}$~g~s$^{-1}$, with a wind duration lasting from ten minutes to one hr using equation 13 and 14 from \citet{Yu2018ApJ...863...53Y}, respectively. This corresponds to a total wind mass loss of (0.35--3.5)$\times$10$^{22}$~g. For a maximum accretion luminosity of 10$^{38}$~erg~s$^{-1}$ in a high state of \source  \citep{Mondal2016MNRAS.461.1917M, Marino2023MNRAS.525.2366M}, the mass accretion rate can be calculated to be 6$\times$10$^{17}$~g~s$^{-1}$, which is at least ten times lower than the anticipated wind mass loss rate during the superburst, depending on the ignition depth. At these high wind mass loss rates, the accretion flow can be effectively suppressed and contaminated with metal-rich nuclear ashes from the burst wind, as observed in our study. The evidence of disrupted accretion flow during the 1999 superburst has also been reported  \citep{Ballantyne2004ApJ...602L.105B}.  

Alternatively, if the accretion flow composition remained unaltered, the absorption lines could instead originate from the expanding burst wind ejecta of the long burst. In this scenario, the wind ejecta, enriched with heavy elements, could produce discrete absorption features. The observed decrease in line energies would then be due to a reduction in the ionization state of the ejected material as it transitions from highly ionized to a neutral state. The expanding ejecta scenario may not be supported due to the measured relatively low value of absorption column density, close to the Galactic value, during the aftermath phase. Moreover, expanding burst ejecta are less likely to remain strong enough during the post-superburst phase, given the typically shorter duration ($<$1~hr) of wind outflows.

We suggest that the observed absorption lines during the aftermath emission most likely originate from a metal-enriched accretion flow. Their detection is intrinsically linked to the nuclear ashes formed during thermonuclear burning on the NS surface. This phenomenon is similar to the spectral features observed from ejected nuclear ashes during short PRE bursts. However, it differs in form due to the significant enrichment of the NS environment with nuclear ashes, driven by wind mass loss during the superburst.

\subsection{Identification of absorption features from the aftermath emission}

Following the recovery phase, the initial line energies at 1.61, 2.75, and 4.0 keV gradually shifted to 1.57, 2.64, and 3.64 keV, respectively. This change in line energy is likely due to the NS's gravitational redshift, rather than decreasing ionization, considering that the accretion flow heats up during the aftermath emission. We have also discussed the aspect of rotational broadening (see Section~\ref{rot}), demonstrating that these lines are not broad enough to support their origin from the NS surface.

Given the similarity of the expected composition of ejected nuclear ashes for the regular bursts and the superburst, the absorption lines observed during the aftermath, with initial energies around 1.61, 2.75, and 4.0 keV, can be tentatively attributed to K$\alpha$ transitions of $^{28}$Si, $^{36}$Ar, $^{40}$Ca, or $^{44}$Ti. In their neutral states, these elements have K$\alpha$ rest-frame energies of 1.74, 2.96, 3.69, and 4.51 keV, respectively. The observed energies at 1.61 and 2.75~keV are well below the neutral transitions for Si and Ar. The 4.0 keV feature may correspond to either neutral Ti, with a K$\alpha$ energy of 4.51 keV, or H-like Ca, whose K$\alpha$ energy near 4.1 keV closely matches the observed value. If the 4.0 keV feature originates from H-like Ca, the remaining lines are likely from H-like states as well, with Si and Ar showing rest-frame K$\alpha$ transitions at 2.0 and 3.32 keV, respectively. Therefore, like the short X-ray bursts, we propose that the observed absorption lines from the aftermath also show the element signatures of Si, Ar, Ca, or Ti.

\subsection{Emission lines}

In our study, we observed the 1 keV emission line with energies ranging from 0.99 to 1.06~keV during the PRE and decay phases of X-ray bursts from \source. Moreover, the 1~keV emission line energy remained almost stable (0.96--1.02~keV) during the aftermath emission. A 1~keV emission feature has been observed during X-ray bursts from various sources, including \source (see, e.g., \citealt{Degenaar2013ApJ...767L..37D, Bult2019ApJ...885L...1B, Strohmayer2019ApJ...878L..27S, Bult2021ApJ...920...59B}). For example, a 1~keV line was detected during the cooling tail of a long burst from IGR~J17062--6143 \citep{Degenaar2013ApJ...767L..37D}. This feature was suggested to originate from Fe~L-band transitions due to the reprocessing of burst photons by the cold gas in the accretion disk. The authors also noted a slight shift in the line energy, which might result from varying charge states during the X-ray burst. The Ly$\alpha$ transition of Ne~X can also produce a line at 1.022 keV. This is possible given that some ultra-compact X-ray binaries exhibit Ne overabundances (see e.g., \citealt{Degenaar2013ApJ...767L..37D, Strohmayer2019ApJ...878L..27S}). 

Given the absence of any significant change in the 1~keV line energy during both the X-ray bursts and the aftermath emission of the long burst, we suggest that this feature is likely due to the reflection of photons by colder gas in the accretion disk. Similarly, the additional emission line observed around 0.7--0.8~keV in the aftermath emission can originate from iron L$_\alpha$ and L$_\beta$ transitions.

\subsection{Impact of rotational broadening on surface absorption lines}  \label{rot}

 The absorption lines detected during the short X-ray bursts and in the aftermath emission are generally narrow. 
However, maximum line widths of 0.3~keV are observed for two absorption features detected around 3 and 4 keV during the aftermath emission (Table~\ref{tab:aftermath}). 
This corresponds to a $\Delta$E/E = $\Delta$$\lambda$/$\lambda$ to be 0.1 and 0.075, respectively. Using Equation 1 from \citet{Lin2010ApJ...723.1053L}, we can anticipate the maximum rotational broadening $\Delta$$\lambda$/$\lambda$ of a line to be 0.3, assuming the NS's spin frequency of 716~Hz, a radius of 10 km, and a 90$^{\circ}$ inclination angle between the NS's rotation axis and the observer line of sight for \source.  This suggests a maximum line width of $\approx$1.2~keV for the 4 keV feature originating from the NS's surface elements. 

In an exceptional case, our observed line width can match the anticipated surface broadening if we assume the inclination angle between the NS's rotation axis and the observer's line of sight is within 20$^{\circ}$.  
However, this is lower than the suggested inclination angle of the system which is thought to range from 30$^{\circ}$ to 50$^{\circ}$ \citep{Anderson1997ApJ...482L..69A, Rogantini2021A&A...645A..98R}. This indicates the absorption lines observed during the aftermath and from the short X-ray bursts are unlikely to originate from the NS surface.

\section{Conclusion}

We studied 15 thermonuclear X-ray bursts from \source using \nicer data obtained between 2017 and 2021. Most of these bursts exhibited narrow spectral features around the PRE and/or the decay phases of the bursts. We detected an emission line around 1 keV, and three other absorption lines with energies around 1.7, 3.0, and 3.75 keV in these burst spectra.  Moreover, from co-added PRE spectra of eleven bursts,  we also detected averaged energies for these emission and absorption lines at 1.02$\pm$0.01, 1.61$\pm$0.01, 2.94$\pm$0.01, and 3.74$\pm$0.03~keV, confirming the presence of these features in the X-ray bursts. We tentatively identified the absorption lines around 1.7, 3.0, and 3.75~keV as likely originating respectively from H-like Si, Ar, and Ca heavy nuclear ashes synthesized during thermonuclear burning. Furthermore, a combined effect of Doppler shift and gravitational redshift is expected from the burst wind, due to a slight shift in observed absorption line energies relative to the rest-frame values of these proposed species.

We also analyzed the aftermath emission of a long burst in 2021 August, which was recovering from a flux depression likely caused by the long burst (superburst). During the aftermath emission, we detected two emission lines (0.7 and 1.0~keV) and three absorption lines. The energies of these absorption lines gradually decreased during the recovery period. Following previous studies of accretion emission from \source and the nature of the white dwarf companion in ultra-compact X-ray binaries, we propose that the observed absorption features may originate from the accretion flow, but only if the flow is significantly contaminated by nuclear ashes from the superburst. This contamination is temporary, where a significant amount of mass loss during the PRE phase of the superburst enhances the accretion flow metallicity.  We propose that the observed absorption lines originate from the same type of elements found in the nuclear ashes. Therefore, irrespective of whether observed during the short X-ray bursts from the direct radiation-driven burst wind or during the aftermath of the superburst from the ash-contaminated accretion flow, the absorption features in both cases would originate from heavy element ashes formed by nucleosynthesis reactions on the NS surface.

On the other hand, the energy of the 1~keV emission line remained nearly constant during both the aftermath and the X-ray bursts. We identified the 1~keV emission line as originating from the Fe~L-band or Ne~X, due to the reflection of the photons off the cold gas in the accretion disk.

\section*{Acknowledgements}
We thank the reviewer for constructive suggestions that improved the manuscript. We also thank Jon Miller for helpful discussions. This work was supported by NASA through the \nicer mission and the Astrophysics Explorers Program and made use of data and software provided by the High Energy Astrophysics Science Archive Research Center (HEASARC). HS is supported by the US NSF under award PHY-2209429. This work has benefitted from interactions facilitated by the International Research Network for Nuclear Astrophysics (IReNA) with support from NSF award OISE-1927130 and the Center for Nuclear Astrophysics across Messengers (CeNAM) with support from the US Department of Energy, Office of Science, Office of Nuclear Physics, under Award Number DE-SC0023128.

\bibliography{references}{}
\bibliographystyle{aasjournal}


\appendix
\restartappendixnumbering
\renewcommand{\thetable}{A\arabic{table}}

\section{X-ray burst quenching  - an effect of superburst? } \label{discuss-superbursteffect}

In this section, we present additional evidence and further discuss the likelihood that the long burst observed in 2021 August \citep{Serino2021ATel14871....1S} was a superburst. \source was monitored by \nicer and INTEGRAL while in a low-hard state in 2021 October, approximately six months after its previous low-hard state in 2021 April-May, during which \nicer detected Burst~\#15 (see Figure~\ref{maxi-bat-lc}). From 2021 October 13 to November 11, we observed the source with \nicer, accumulating a total exposure time of 75~ks, but no X-ray bursts were detected in these data sets.
Furthermore, INTEGRAL/JEM-X \citep{Lund2003A&A...411L.231L} conducted observations between 2021 October 17 and 18  (MJD~59504.47 -- 59505.13), around the minima of the low-hard state observed with MAXI (Figure~\ref{maxi-bat-lc}). We created 3--25~keV light curves with JEM-X1 and JEM-X2 at 1s time resolution using the OSA~11.2 analysis software. Despite the nearly uninterrupted exposure time of 54~ks, no X-ray burst was found with INTEGRAL/JEM-X. 
Given that \source typically shows X-ray bursts with a recurrence time of 2–-4 hours in its low-hard state \citep{Grindlay1976ApJ...205L.127G, Chou2001ApJ...563..934C, Galloway2008ApJS..179..360G, Zand2012A&A...547A..47I}, we would have expected to detect about a dozen X-ray bursts during the INTEGRAL observations. However, the non-detection of X-ray bursts with both \nicer and JEM-X in 2021 October suggests that bursting activity was still quenched after the candidate superburst on 2021 August 23. Stable burning of the material may keep the surface hot in the quenching period that may last for weeks to months \citep{Kuulkers2004NuPhS.132..466K, Keek2012ApJ...752..150K}. Thus, it is most plausible that the long burst observed in 2021 August was indeed a superburst from \source.

MAXI reported another long burst from the globular cluster NGC~6624 on 2021 November 25 (MJD 59543.975) \citep{Serino2021ATel15071....1S}. Surprisingly, the burst was observed more than 15 days after the end of the source's low-hard state in 2021 October (see Figure~\ref{maxi-bat-lc}). The second long burst might be an intermediate-duration burst from \source, as it is unlikely for two superbursts to occur from the same source within a three-month period \citep{Cumming2003ApJ...595.1077C}. Intermediate-duration bursts can last several tens of minutes and are caused by the ignition of a thick helium layer slowly accumulated on the NS \citep{Zand2005A&A...441..675I, Cumming2006ApJ...646..429C, Alizai2023MNRAS.521.3608A}. The accumulation of material on the NS surface during and after the quenching period may provide sufficient fuel for a long burst.  
Alternatively, the 2021 November long burst might have originated from another source within the metal-rich globular cluster NGC~6624, which hosts multiple white dwarfs and six radio pulsars in addition to the ultra-compact X-ray binary \source \citep{Revnivtsev2002AstL...28..237R}.

Regarding the superburst recurrence, \source is predicted to produce a superburst approximately every 5 to 10 years if the accreted material is primarily helium with a low hydrogen abundance \citep{Cumming2003ApJ...595.1077C}. The expected recurrence shortens to about 1 to 2 years for pure helium accretion \citep{Cumming2003ApJ...595.1077C}. Observationally, MAXI and ASM/\rxte detected a candidate superburst from NGC~6624 in 2010 March \citep{Zand2011ATel.3625....1I}, which occurred roughly 10.5 years after a superburst observed by \rxte in 1999 September \citep{Strohmayer1997ApJ...487L..77S}. Another candidate superburst (long burst) was detected by MAXI in 2021 August \citep{Serino2021ATel14871....1S}, about 11.45 years after the 2010 event. These intervals support the hypothesis of a decade-long recurrence period for superbursts from \source (see \citealt{Strohmayer2002ApJ...566.1045S}), assuming no superburst event was missed by MAXI or other X-ray sky monitoring instruments.

\end{document}